\documentclass[english,letter,onecolumn]{IEEEtran}
\usepackage[T1]{fontenc}
\usepackage[latin9]{inputenc}
\usepackage{amsthm}
\usepackage{amsmath}
\usepackage{amssymb}
\usepackage{graphicx}
\usepackage{color}

\makeatletter
\theoremstyle{plain}
\newtheorem{thm}{\protect\theoremname}
\theoremstyle{definition}
\newtheorem{defn}{\protect\definitionname}
\theoremstyle{plain}
\newtheorem{lem}{\protect\lemmaname}

\usepackage{babel}

\makeatother

\usepackage{babel}
\providecommand{\definitionname}{Definition}
\providecommand{\lemmaname}{Lemma}
\providecommand{\theoremname}{Theorem}

\begin{document}

\title{Channel Diversity needed for Vector Space Interference Alignment}

\author{Cheuk Ting Li and Ayfer \"{O}zg\"{u}r\\
 % Department of Electrical Engineering\\
 Stanford University, California, USA\\
 Email: ctli@stanford.edu, aozgur@stanford.edu\\
\thanks{The work of C. T. Li was supported in part by a Hong
Kong Alumni Stanford Graduate Fellowship. The work of A. Özgür was supported in part by NSF CAREER award 1254786 and by the Center for Science of Information (CSoI), an NSF Science and Technology Center, under grant agreement CCF-0939370. 
This paper was presented in part at the IEEE International
Symposium on Information Theory, Honolulu, USA, June 2014.%

This paper is published in IEEE Transactions on Information Theory (Volume: 62, Issue: 4, April 2016), available at http://ieeexplore.ieee.org/document/7406708/ .
Copyright (c) 2014 IEEE. Personal use of this material is permitted.  However, permission to use this material for any other purposes must be obtained from the IEEE by sending a request to pubs-permissions@ieee.org.
}}
\maketitle
\begin{abstract}
We consider vector space interference alignment strategies
over the $K$-user interference channel and derive an upper bound
on the achievable degrees of freedom as a function of the channel
diversity $L$, where the channel diversity  is modeled by $L$  real-valued parallel channels with coefficients  drawn from a non-degenerate joint distribution. The seminal work of Cadambe and Jafar shows that when $L$ is unbounded, vector space interference alignment can achieve $1/2$ degrees of freedom per user independent of the number of users $K$. However wireless channels have limited diversity in practice, dictated by their coherence time and bandwidth, and an important question is the number of degrees of freedom achievable at finite $L$. When $K=3$ and if $L$ is finite, Bresler et al show that the number of degrees of freedom achievable with vector space interference alignment is bounded away from  $1/2$, and the gap decreases inversely proportional to $L$. In
this paper, we show that when $K\geq4$, the gap is significantly larger. In particular, the gap to the optimal $1/2$ degrees of freedom per user can decrease at most like $1/\sqrt{L}$,
and when $L$ is smaller than the order of $2^{(K-2)(K-3)}$, it decays at most like $1/\sqrt[4]{L}$.
\end{abstract}
\begin{keywords} Interference alignment, K-user interference channel, degrees of freedom, channel diversity, blocklength.
\end{keywords}

\section{Introduction}

Interference is the central phenomenon severely limiting the performance
of most wireless systems. Over the recent years, interference alignment
has emerged as a promising tool to mitigate interference \cite{maddahali2008,Cadambe--Jafar2008}.
The main idea is to design transmit signals of different users in
such a way that, upon arriving at the unintended receivers, they overlap
with each other and the resulting interference is perceived as much
less than the sum of the individual interferences. Surprisingly, the
work \cite{Cadambe--Jafar2008} of Cadambe and Jafar has shown that
this approach can lead to $K/2$ sum degrees of freedom over the time
or frequency-varying $K$-user interference channel, while traditional
approaches such as treating interference as noise or orthogonalizing
transmissions can provide only one degree of freedom. This roughly
implies that at high-SNR, each user can communicate as if it has half
the resources of the channel for its exclusive use, regardless of
the total number of users.

However, one of the main caveats of the $K/2$ degrees of freedom
result in \cite{Cadambe--Jafar2008} is that it requires unbounded time or frequency
variation of the channel. More precisely, in order to achieve $K/2$ degrees of freedom, the transmitters have to code over the order of $K^{K^{2}}$ independent realizations of the channel (or equivalently $K^{K^{2}}$ parallel channels). (This scaling is slightly improved to $2^{K^{2}}$ by Özgür and Tse~\cite{OzgurTse08}.) In practice, wireless channels have finite
channel diversity dictated by their coherence time and bandwidth, and the requirement $K^{K^{2}}$ is prohibitive even for small values of $K$. Whether this exponential requirement for channel diversity is fundamental or not to vector space interference
alignment strategies, of which the scheme in \cite{Cadambe--Jafar2008}
is one specific example, is an important question in determining
the real potential of interference alignment in practical wireless systems.

Despite significant research interest in interference alignment
 over the recent years (see \cite{JafarNow} for an overview),
there is limited understanding regarding this question, and more generally, regarding how the available channel diversity impacts the ability to align interference. The problem is understood only in the case when $K=3$. In this case, Bresler and Tse~\cite{bresler2009} characterize the exact relation between the channel diversity $L$, modeled by the number of independent channel
realizations over time or frequency, and the total number of degrees of freedom achievable using vector space alignment. (Their
result subsumes an earlier result by Cadambe, Jafar and Wang~\cite{CJW08} which corresponds
to the special case $L=2$.) They show that the achievable sum degrees
of freedom in the $3$-user interference channel are given by 
\begin{equation}\label{eq:Bresler}
\mathrm{DoF}=\frac{3}{2}\left(1-\frac{1}{4L-2\lfloor L/2\rfloor-1}\right).
\end{equation}
We can observe that when $L\rightarrow\infty$, $3/2$ degrees of
freedom are achievable as expected, and for finite values of $L$
the formula precisely characterizes how $\mathrm{DoF}$ approaches
$3/2$ as a function of $L$. To our knowledge, nothing is
known regarding the relation between channel diversity and achievable
degrees of freedom for interference channels with more than $3$-users;
apart from the trivial conclusion that when $L=1$, vector interference
alignment can achieve only one degree of freedom and the result of
\cite{Cadambe--Jafar2008} which shows that when $L\rightarrow\infty$,
$K/2$ degrees of freedom are achievable.

In this paper, which is an extended and more complete version of~\cite{li2014channel}, we make progress in this direction by first showing
that for $K\geq4$, 
\[
\mathrm{DoF}\le\frac{K}{2}\left(1-\frac{1}{11\sqrt{L}}\right).
\]
This result shows that the degrees of freedom per user approach $1/2$
at a much slower speed when $K\geq4$ when compared to $K=3$: the
gap decreases at most like $1/\sqrt{L}$ as opposed to $1/L$. Next,
we further improve our result to 
\[
\mathrm{DoF}\le\frac{K}{2}\left(1-C\min\left\{ \frac{1}{\sqrt[4]{L}},\frac{2^{\left(K-2\right)\left(K-3\right)/4}}{\sqrt{L}}\right\} \right),
\]
where $C>0$ is a constant. In the regime when $L$ is smaller than
the order of $2^{\left(K-2\right)\left(K-3\right)}$, i.e., when the first term of the minimum
is smaller than the second, this implies that the gap to the optimal
$1/2$ degrees of freedom per user decreases at most like $1/\sqrt[4]{L}$.
As a result, when $K$ grows, either we need an exponential channel diversity $L > 2^{\left(K-2\right)\left(K-3\right)}$, or the gap to the optimal 1/2 degrees of freedom per user decreases at most like $1/\sqrt[4]{L}$.
%Since an exponential channel diversity is impractical for moderately sized $K$, the gap decreases at most like $1/\sqrt[4]{L}$ in practical scenarios.

A closer look at the scheme in \cite{Cadambe--Jafar2008} reveals
that the following degrees of freedom are achievable over the $K$-user
interference channel for $L$ large enough. 
\begin{equation}\label{eq:ach}
\mathrm{DoF}\geq\frac{K}{2}\left(1-\frac{CN}{\sqrt[N]{L/2}}\right),
\end{equation}
where $N=(K-1)(K-2)-1$ and $C>0$ is a constant. When $K=3$, we have $N=1$ and this matches the scaling in \eqref{eq:Bresler}. When $K=4$, we have $N=5$ which implies that gap to the optimal degrees of freedom decreases like $1/\sqrt[5]{L}$ in \eqref{eq:ach}, while our upper bound only implies that the gap can not decrease faster than $1/\sqrt{L}$ ($1/\sqrt[4]{L}$ when $L$ is smaller than the order of $2^{\left(K-2\right)\left(K-3\right)}$). The difference between the scaling of our upper bound and the achievability in \eqref{eq:ach} becomes even larger as $K$ increases.

While the remaining gap between the lower bound \eqref{eq:ach} and the upper bounds we derive is still quite large, one of the main contributions of this paper is to build a mathematical framework (tools and notions) for studying the alignment problem when $K\geq 4$. Note that the case $K\geq 4$ is significantly more complex than the case $K=3$, in which case it is possible to explicitly keep track of how intertwined the users' signaling strategies are due to alignment. The exact characterization in \eqref{eq:Bresler} is indeed based on such explicit tracking of users' signaling spaces. For $K\geq4$, there is significantly more freedom in choosing user's signaling spaces and it is not possible to keep track of the intertwining between them. Without such explicit tracking, we provide a framework that allows to capture the tradeoff between the two requirements of aligning interference at the unintended receivers and that of keeping the desired signal space distinct from interference at the intended receivers. We believe this framework can be further developed to prove tighter results in the future.

\subsection{Related Work}

A related problem has been considered in a recent paper \cite{sun2015interference}, which restricts each transmitter to send a single beam (the signaling space of each transmitter has dimension one) and asks how many transmitter-receiver pairs can be accommodated when the channel diversity is finite. Their approach combines counting arguments with algebraic tools to determine the feasibility of a hybrid system of equations and inequalities. In contrast here we do not restrict the dimension of the signaling space at each transmitter. Indeed,  \cite{Cadambe--Jafar2008} shows that the benefits of the interference alignment are asymptotic in nature and can be realized by increasing the dimension of the signaling space at the transmitters, which leads to more freedom in the choice of the signaling spaces. This, however, also makes the problem of characterizing the achievable degrees of freedom more difficult and in particular one can not rely on explicit counting arguments as in \cite{sun2015interference}.

Another related line of research \cite{gomadam2011,Yetis,bresler2011,Razaviyayn}
(see also \cite{JafarNow} and the references therein) looks at the
relation between the spatial diversity available in a MIMO interference
channel and the degrees of freedom achievable with vector interference
alignment strategies. Here each user is equipped with multiple antennas
and signals are aligned over the spatial dimension with no time/frequency
diversity in the channel. The impact of the spatial diversity (number of transmit and receive antennas) on the
achievable degrees of freedom with vector interference alignment strategies
is much better understood. For example, \cite{bresler2014} shows
that in the symmetric case where each node is equipped with $N$ antennas,
the maximum number of $\mathrm{DoF}$ achievable with vector space
alignment strategies is given by 
\[
\mathrm{DoF}=K\left\lfloor \frac{2N}{K+1}\right\rfloor \leq2N\frac{K}{K+1}.
\]
In sharp contrast to the $K/2$ degrees of freedom achievable with
time/frequency diversity, this result implies that the $\mathrm{DoF}$
gain from aligning interference over the spatial dimension is limited
by a factor of $2$ when compared to the $\mathrm{DoF}$ achieved
with simple orthogonalization of users' transmissions. This
implies that the gain from spatial interference alignment is very
limited when compared to the potential gain from aligning interference
over time/frequency varying channels. Therefore, we believe understanding
the feasibility of interference alignment over time/frequency varying
channels with limited diversity is the key to assessing the real potential
of interference alignment strategies in practical systems. %The fully-connected $K$-user interference channel is of central importance in wireless communication. When more users are present, the intended signal, being obfuscated by more interference, would be more difficult to decode. It is rather surprising that using a scheme termed interference alignment \cite{maddahali2006,jafar2008,Cadambe--Jafar2008}, the interference would only reduce the degree of freedom per user by a half. Cadambe and Jafar \cite{Cadambe--Jafar2008} proved that $K/2$ degrees of freedom is achievable in the $K$-user interference channel using interference alignment, attaining the outer bound given by H{\o}st-Madsen and Nosratinia \cite{hostmadsen2005}.

\section{Problem Formulation}

%\textbf{introduce the channel model, what we mean by vector interference alignment and degrees of freedom, I think we need a little bit more detail here, please check Bresler's thesis and paper} 

\subsection{Notation}

For a vector $\mathbf{v}\in\mathbb{R}^{L}$, we write $\left\Vert \mathbf{v}\right\Vert _{0}$
for the number of nonzero entries of $\mathbf{v}$. For $\mathbf{H}\in\mathbb{R}^{L\times L}$ and subspace $V \subseteq \mathbb{R}^{L}$, we write $\mathbf{H}V$ for the subspace $\{\mathbf{H}\mathbf{v} \,:\, \mathbf{v} \in V\}$. For subspaces
$V_{1},V_{2},...,V_{n}\subseteq\mathbb{R}^{L}$, we write $V_{1}+V_{2}=\mathrm{span}\left(V_{1}\cup V_{2}\right)$,
and $\sum_{i=1}^{n}V_{i}=V_{1}+\cdots+V_{n}$. We write $\left\langle \mathbf{v}_{1},...,\mathbf{v}_{n}\right\rangle =\mathrm{span}\left\{ \mathbf{v}_{1},...,\mathbf{v}_{n}\right\} $.
For a subset $S\subseteq\left\{ 1,...,L\right\} $, $\mathbb{R}^{S}=\{\mathbf{v} \in \mathbb{R}^L:\, \mathbf{v}_i = 0 \;\text{for all}\; i \notin S\}$. The $L\times L$ identity matrix is denoted by $\mathbf{I}_L$ ($L$ may be omitted when the dimension is clear in the context). For a vector $\mathbf{v}\in\mathbb{R}^{L}$, we
write $\mathrm{diag}(\mathbf{v})\in\mathbb{R}^{L\times L}$ for the
diagonal matrix formed by the entries of $\mathbf{v}$. For $\mathbf{X}\in\mathbb{R}^{L\times L}$,
we write $\mathrm{diag}(\mathbf{X})\in\mathbb{R}^{L}$ for the vector
formed by the diagonal entries of $\mathbf{X}$.

\subsection{Channel Model}

Consider the fully-connected $K$-user Gaussian interference channel,
where receiver $i$ wants to obtain a message from transmitter $i$
for $1\leq i\leq K$, but the signal received is superimposed by interferences
from transmitters $j\neq i$. The input-output relationship is given
by 
\begin{equation}\label{eq:interferencech}
\mathbf{y}_{i}=\sum_{j=1}^{K}\mathbf{H}_{ij}\mathbf{x}_{j}+\mathbf{z}_{i},
\end{equation}
where $\mathbf{x}_{i}\in\mathbb{R}^{L}$ is the transmitted signal
of transmitter $i$ over $L$ channel uses; $\mathbf{y}_{i}\in\mathbb{R}^{L}$
is the received signal of receiver $i$; $\mathbf{z}_{i}\sim\mathcal{N}\left(0,\mathbf{I}\right)$
is an additive white Gaussian noise; and $\mathbf{H}_{ij}\in\mathbb{R}^{L\times L}$
is a diagonal matrix containing the channel coefficients from Transmitter
$j$ to Receiver $i$ over the $L$ channel uses, 
\[
\mathbf{H}_{ij}=\left[\begin{array}{ccc}
h_{ij}^{(1)}\\
 & \ddots\\
 &  & h_{ij}^{(L)}
\end{array}\right].
\]
We assume the entries of $\mathbf{H}_{ij}$ are chosen i.i.d. from
a continuous distribution, or more generally, the joint distribution
of $\left\{ (\mathbf{H}_{ij})_{\ell\ell}\right\} _{i,j=1,...,K,\,\ell=1,...,L}$
has a density in the $LK^{2}$-dimensional space. This channel model
corresponds to $L$ uses of a fast fading interference channel where
we get a different realization of the channel at each use.

The integer $L$ is called the \emph{diversity} of the channel. In
the above model it is related to the blocklength of communication, and more precisely, it is the number of coherence periods over which we
code.
%\textcolor{red}{For the block fading case where each coherence period is of
%duration $T$, the matrices $\mathbf{H}_{ij}$ are formed by placing
%$T$ copies of $\mathrm{diag}[h_{ij}^{(1)},...,h_{ij}^{(L)}]$ diagonally,
%i.e., $\mathbf{H}_{ij}=\mathbf{I}_{T}\otimes\mathrm{diag}\left[h_{ij}^{(1)},...,h_{ij}^{(L)}\right]\in\mathbb{R}^{TL\times TL}$,
%where $\otimes$ denotes the Kronecker product.}
For the block fading case where each coherence period is of
duration $T$, $\mathbf{H}_{ij}$ are the diagonal matrices formed by
$h_{ij}^{(1)},\ldots,h_{ij}^{(1)},h_{ij}^{(2)},\ldots,h_{ij}^{(2)},h_{ij}^{(3)},\ldots,h_{ij}^{(L)}$, where each $h_{ij}^{(l)}$ is repeated $T$ times,
i.e., $\mathbf{H}_{ij}=\mathrm{diag}\left[h_{ij}^{(1)},...,h_{ij}^{(L)}\right]\otimes \mathbf{I}_{T} \in\mathbb{R}^{TL\times TL}$,
where $\otimes$ denotes the Kronecker product.
In this paper, we
first consider the fast fading case ($T=1$) and then extend our results
to the block fading case.

\subsection{Vector Interference Alignment Strategies and Degrees of Freedom}

In this paper we focus on vector space schemes, which we specify next.
Suppose transmitter $i$ wishes to transmit $\widehat{\mathbf{x}}_{i}\in\mathbb{R}^{D}$
containing $D$ data symbols. It applies a precoding matrix $\mathbf{V}_{i} \in \mathbb{R}^{L \times D}$
and transmits $\mathbf{x}_{i}=\mathbf{V}_{i}\widehat{\mathbf{x}}_{i}$.
Let $V_{i}\subseteq\mathbb{R}^{L}$ be the column span of $\mathbf{V}_{i}$.
Receiver $i$ decodes $\widehat{\mathbf{x}}_{i}$ by zero-forcing interference,
i.e., projecting its received signal on the orthogonal complement of
the space spanned by the interference. At high SNR, it can decode
the $D$ data symbols if the signal subspace $\mathbf{H}_{ii}V_{i}$
intersects the interference subspace only at 0, i.e., 
\[
\mathbf{H}_{ii}V_{i}\cap\biggl(\sum_{j\neq i}\mathbf{H}_{ij}V_{j}\biggr)=\left\{ 0\right\} .
\]
We call this the \emph{decoding condition} at receiver $i$. The maximum
total degrees of freedom achievable by this strategy is given by 
\[
\mathrm{DoF}=\max_{\left\{ V_{i}\right\} \text{ satisfies decoding condition }\forall i}KD/L.
\]
It is easy to observe that this corresponds to the classical degrees
of freedom definition for the interference channel: In particular assume that the transmitted signals $\mathbf{x}_{i}\in\mathbb{R}^{L}$ in \eqref{eq:interferencech} are subject to an average power constraint $LP$, i.e. average power $P$ per channel use.  The total degrees of freedom achieved by the vector interference alignment strategy can be equivalently defined as
$$
\mathrm{DoF}=\lim_{P\to\infty}\frac{1}{L}\frac{R(P)}{\log P}
$$
where $R(P)$ denotes the rate achieved by this strategy under  a per user power constraint $P$.

If we wish to have $\mathrm{DoF}\ge\left(1-\epsilon\right)K/2$, then
$D\geq\left(1-\epsilon\right)L/2$. Given that the signalling subspaces $V_i$ have to satisy the decoding condition at each receiver, the goal of this paper is to give
a lower bound on the channel diversity $L$ in terms of the gap $\epsilon$. This translates
to an upper bound on the achievable degrees of freedom with any given
channel diversity $L$.

In the block fading case, the signal space is $V_{i}\subseteq\mathbb{R}^{TL}$
instead of $V_{i}\subseteq\mathbb{R}^{L}$, and therefore the definition
of maximum total degrees of freedom is modified as 
\[
\mathrm{DoF}=\max_{\left\{ V_{i}\right\} \text{ satisfies decoding condition }\forall i}\frac{KD}{TL}.
\]

\section{Main Result}

The following theorem is the main result of this paper. 
\begin{thm}
\label{thm:prob_quad}In the fast fading case ($T=1$), when $K\ge4$,
with probability $1$, the maximum sum degrees of freedom achievable
with vector space interference alignment strategies is bounded by
\[
\mathrm{DoF}\le\frac{K}{2}\left(1-\frac{1}{11\sqrt{L}}\right).
\]

\end{thm}
The theorem can be extended to block fading, at the expense of a larger
constant. 
\begin{thm}
\label{thm:prob_quad_block}In the block fading case for any value
of $T\ge1$, when $K\ge4$, with probability $1$, the maximum sum
degrees of freedom achievable with vector space interference alignment
strategies is bounded by 
\[
\mathrm{DoF}\le\frac{K}{2}\left(1-\frac{1}{20\sqrt{L}}\right).
\]

\end{thm}
The result can be improved for large $L$ and $K$ to the following
result. 
\begin{thm}
\label{thm:better_bound}In the fast fading or block fading case for
any value of $T\ge1$, when $K\ge4$, with probability $1$, the maximum
total degrees of freedom is bounded by

\[
\mathrm{DoF}\le\frac{K}{2}\left(1-2^{-17}\min\left\{ \frac{1}{\sqrt[4]{L}},\frac{2^{\left(K-2\right)\left(K-3\right)/4}}{\sqrt{L}}\right\} \right).
\]

\end{thm}

Although the constant in this theorem is quite small, we believe the theorem and its proof are important in illustrating how the notions and the tools we develop to tackle this problem (such as extension and contraction of a subspace defined in the next section) can be further developed in nontrivial ways to obtain tighter results.

The rest of the paper is devoted to the proof of the theorems. In
Section~\ref{sec:LinAlgProb}, we define and develop three notions:
the alignment width of a subspace, the sparsity of a subspace, and
the linear independence condition for a set of diagonal matrices which
allow us to convert the problem of interest to a pure linear algebra
problem. In Section~\ref{sec:ProofIntuition}, we provide the intuition
for our proof under a simplifying assumption. The proof of our main
result for fast fading (Theorem~\ref{thm:prob_quad}) is given in
Section \ref{sec:LBDiversity}, and for block fading (Theorem~\ref{thm:prob_quad_block})
in Section \ref{sec:BlockFading}. Theorem \ref{thm:better_bound}
is proved in Section \ref{sec:BetterBound}.

\section{A Linear Algebra problem\label{sec:LinAlgProb}}

Below, we focus on the case $K\ge4$. We assume that the diagonal
entries of $\mathbf{H}_{ij}$ are nonzero, which holds with probability
1.

\subsection{Alignment Width}

\begin{defn}[Extension and contraction operators]
Let $V\subseteq\mathbb{R}^{L}$ be a subspace, and $\mathbf{T}\in\mathbb{R}^{L\times L}$
be a diagonal matrix with non-zero diagonal entries. Define the \emph{extension operator} $\mathrm{e}_{\mathbf{T}}$
and the \emph{contraction operator} $\mathrm{c}_{\mathbf{T}}$ by 
\begin{eqnarray*}
\mathrm{e}_{\mathbf{T}}V & = & V+\mathbf{T}V,\\
\mathrm{e}_{\mathbf{T}}^{n}V & = & V+\mathbf{T}V+\cdots+\mathbf{T}^{n}V,\\
\mathrm{c}_{\mathbf{T}}V & = & V\cap\mathbf{T}V,\\
\mathrm{c}_{\mathbf{T}}^{n}V & = & V\cap\mathbf{T}V\cap\cdots\cap\mathbf{T}^{n}V.
\end{eqnarray*}
\end{defn}

\begin{figure}[t!]
\centering \scalebox{0.25}{\includegraphics{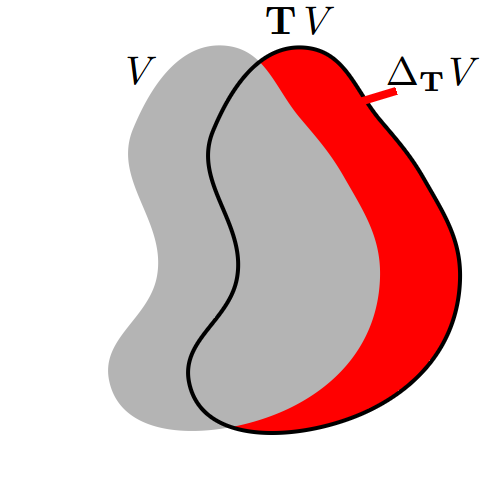}} \caption{Illustration of alignment width of $V$ under $\mathbf{T}$. Multiplication
by $\mathbf{T}$ is represented by a shift to the right.}

\label{fig:alinmentwidth} 
\end{figure}

\begin{defn}[Alignment width]
We define the \emph{alignment width} of a subspace $V$ under a diagonal
matrix $\mathbf{T}$ by 
\begin{eqnarray*}
\Delta_{\mathbf{T}}V & = & \dim\left(\mathrm{e}_{\mathbf{T}}V\right)-\dim V\\
 & = & \dim V-\dim\left(\mathrm{c}_{\mathbf{T}}V\right),
\end{eqnarray*}
\end{defn}
The equality is due to
\begin{align}
\dim(V+W) &= \dim(V)+\dim(W)-\dim(V\cap W) \label{eq:dim_sum_int}
\end{align}
for subspaces $V,W$. This equality will be used extensively throughout the paper.
Intuitively, the alignment width is a measure of the difference
between $V$ and its rotated version $\mathbf{T}V$; it is the dimension
of the subspace which jumps out of the original subspace after the
linear transformation by $\mathbf{T}$. Equivalently, according to the second equivalent definition it can be thought of as the dimension of the part of $V$ that does not align with $\mathbf{T}V$. This is illustrated in Figure~\ref{fig:alinmentwidth}.

There are several properties of extension and contraction operators that follow directly from their definitions and will be used repeatedly throughout the paper. Extensions along different matrices commute with each other, and so do contractions, i.e.,
\begin{align*}
\mathrm{e}_{\mathbf{T}_1}\mathrm{e}_{\mathbf{T}_2}V &= \mathrm{e}_{\mathbf{T}_2}\mathrm{e}_{\mathbf{T}_1}V, \\
\mathrm{c}_{\mathbf{T}_1}\mathrm{c}_{\mathbf{T}_2}V &= \mathrm{c}_{\mathbf{T}_2}\mathrm{c}_{\mathbf{T}_1}V.
\end{align*}
However, extension and contraction do not commute with each other. Instead the following holds
\begin{align}
\mathrm{e}_{\mathbf{T}_1}\mathrm{c}_{\mathbf{T}_2}V &\subseteq \mathrm{c}_{\mathbf{T}_2}\mathrm{e}_{\mathbf{T}_1}V. \label{eq:ec_contain}
\end{align}
Moreover,
\begin{align}
\mathrm{e}_{\mathbf{T}}\mathrm{c}_{\mathbf{T}^{-1}}V &\subseteq V \subseteq \mathrm{c}_{\mathbf{T}^{-1}}\mathrm{e}_{\mathbf{T}}V. \label{eq:ec_contain_eq}
\end{align}

Now, define 
\[
\mathbf{T}_{ijk}=\mathbf{H}_{1i}^{-1}\mathbf{H}_{1k}\mathbf{H}_{jk}^{-1}\mathbf{H}_{ji}.
\]
Since the matrices $\mathbf{H}_{ij}$ are drawn from a continuous distribution, the matrix $\mathbf{T}_{ijk}$ is almost surely defined and invertible, and hence we assume this throughout the paper.
In the following lemma, we show that if the subspaces $V_{i}$ satisfy
the decoding condition, then they have to ``align'' with these diagonal
matrices $\mathbf{T}_{ijk}$ in the sense that $V_{i}$ has a large
intersection with $\mathbf{T}_{ijk}V_{i}$, i.e., $\Delta_{\mathbf{T}_{ijk}}V_{i}$
is small. The lemma builds on the observation that if two signal subspaces
$V_{i}$ and $V_{k}$ have nearly the same projections at two receivers where they consitute interference say Receiver $1$ and Receiver $j$, then $\mathbf{H}_{1i}V_i \approx \mathbf{H}_{1k}V_k$ and $\mathbf{H}_{ji}V_i \approx \mathbf{H}_{jk}V_k$. Hence
\[
\mathbf{H}_{1i}^{-1}\mathbf{H}_{1k}\mathbf{H}_{jk}^{-1}\mathbf{H}_{ji} V_i \approx \mathbf{H}_{1i}^{-1}\mathbf{H}_{1k} V_j \approx V_i.
\]

\begin{lem}[Width requirement for decoding]
\label{lem:wreq} If $D=\left(1-\epsilon\right)L/2$ and $V_{i},\, i=1,\dots,K$
satisfy the decoding condition at all the receivers, then $\Delta_{\mathbf{T}_{ijk}}V_{i}\le2\epsilon L$
for all distinct $i,j,k\neq1$.\end{lem}
\begin{IEEEproof}
Due to the decoding condition at receiver $1$, for any distinct $i,k\neq1$
we have 
\[
\dim\left(\mathbf{H}_{1i}V_{i}\!+\!\mathbf{H}_{1k}V_{k}\right)=\dim\left(V_{i}\!+\!\mathbf{H}_{1i}^{-1}\mathbf{H}_{1k}V_{k}\right)\le\left(1+\epsilon\right)L/2.
\]
Due to the decoding condition at receiver $j\neq1$, we have 
\[
\dim\left(\mathbf{H}_{ji}V_{i}\!+\!\mathbf{H}_{jk}V_{k}\right)\!=\!\dim\left(\mathbf{T}_{ijk}V_{i}\!+\!\mathbf{H}_{1i}^{-1}\mathbf{H}_{1k}V_{k}\right)\!\le\!\left(1+\epsilon\right)L/2.
\]
for any distinct $i,j,k$. Let $\widehat{V}_{k}=\mathbf{H}_{1i}^{-1}\mathbf{H}_{1k}V_{k}$.
Then by \eqref{eq:dim_sum_int},
\begin{align*}
\lefteqn{\dim\left(V_{i}\cap\widehat{V}_{k}\right)}\\
 & =\dim\left(V_{i}\right)+\dim\left(\widehat{V}_{k}\right)-\dim\left(V_{i}+\widehat{V}_{k}\right)\\
 & \ge2D-\left(1+\epsilon\right)L/2,
\end{align*}
and similarly we have $\dim\left(\mathbf{T}_{ijk}V_{i}\cap\widehat{V}_{k}\right)\geq2D-\left(1+\epsilon\right)L/2$.
Hence again  using\eqref{eq:dim_sum_int}, we have
\begin{eqnarray*}
\lefteqn{\dim\left(V_{i}\cap\mathbf{T}_{ijk}V_{i}\right)}\\
 & \ge & \dim\left(V_{i}\cap\mathbf{T}_{ijk}V_{i}\cap\widehat{V}_{k}\right)\\
  & = & \dim\left((V_{i}\cap\widehat{V}_{k})\cap(\mathbf{T}_{ijk}V_{i}\cap\widehat{V}_{k})\right)\\
 & = & \dim\left(V_{i}\cap\widehat{V}_{k}\right)+\dim\left(\mathbf{T}_{ijk}V_{i}\cap\widehat{V}_{k}\right)\\
 &  & \; -\dim\left((V_{i}\cap\widehat{V}_{k})+(\mathbf{T}_{ijk}V_{i}\cap\widehat{V}_{k})\right)\\
 & \ge & \dim\left(V_{i}\cap\widehat{V}_{k}\right)+\dim\left(\mathbf{T}_{ijk}V_{i}\cap\widehat{V}_{k}\right)-\dim(\widehat{V}_{k})\\
 & \ge & 3D-\left(1+\epsilon\right)L
\end{eqnarray*}
and 
\begin{eqnarray*}
\Delta_{\mathbf{T}_{ijk}}V_{i} & = & D-\dim\left(V_{i}\cap\mathbf{T}_{ijk}V_{i}\right)\\
  & \le & D-\left(3D-\left(1+\epsilon\right)L\right)\\
 & = & 2\epsilon L,
\end{eqnarray*}
%\[
%\Delta_{\mathbf{T}_{ijk}}V_{i}=D-\dim\left(V_{i}\cap\mathbf{T}_{ijk}V_{i}\right)\le D-\left(3D-\left(1+\epsilon\right)L\right)=2\epsilon L,
%\]
which completes the proof of the lemma. 
\end{IEEEproof}
$ $

\subsection{Sparsity of Subspaces}

In this section, we define the sparsity of a subspace and show that
if $V_{i}$ satisfy the decoding condition then they cannot have low
sparsity. 
\begin{defn}
($N$-sparsity) We define the \emph{$N$-sparsity} of a subspace $V\subseteq\mathbb{R}^{L}$
as 
\begin{eqnarray*}
\mathrm{sp}_{N}\left(V\right)\!\! & \!\!=\!\! & \!\!\min\left\{ \left|S\right|\,:\, S\subseteq\left\{ 1,...,L\right\} ,\,\dim\left(V\cap\mathbb{R}^{S}\right)\ge N\right\} \\
 & \!\!=\!\! & \!\!\min\left\{ \max_{\mathbf{v}\in W}\left\Vert \mathbf{v}\right\Vert _{0}\,:\, W\subseteq V,\,\dim\left(W\right)\ge N\right\} .
\end{eqnarray*}
When $N>\dim V$, let $\mathrm{sp}_{N}\left(V\right)=\infty$. 
\end{defn}
The $N$-sparsity of a subspace $V$ quantifies the sparsity of its
sparsest $N$-dimensional subspace. Consider the first definition:
if $\mathrm{sp}_{N}(V)=d$, then there exists an $N$-dimensional
subspace of $V$, call it $W$, which is fully contained in $\mathbb{R}^{S}$
for some $S\subseteq\left\{ 1,...,L\right\} $ such that $|S|=d$,
i.e., $W$ is composed of vectors with all entries other than those
in $S$ equal to zero. (This immediately implies that $\mathrm{sp}_{N}(V)\geq N$.)
Hence, $\max_{\mathbf{v}\in W}\left\Vert \mathbf{v}\right\Vert _{0}\leq d$.
Moreover, $V$ has no $N$-dimensional subspace which is only composed
of vectors with fewer than $d$ non-zero entries. Hence in every subspace of
$V$ of dimension equal to (or larger than) $N$, we can find a vector
with at least $d$ non-zero entries. This establishes the equivalence
of the first definition to the second.
Also it follows from the definition that $\mathrm{sp}_{N}(V)$ is non-decreasing in $N$. This fact will be used throughout the paper.

In the following lemma, we show that if the subspaces $V_{i}$ satisfy
the decoding conditions at all the receivers, then they cannot be
too sparse. The lemma builds on the intuition that if $V_{i}$ contains
a large dimensional sparse subspace then it remains largely unchanged
under the direct link and cross link transformations. This contradicts
the requirement that $V_{i}$ has to align with the other signal
subspaces at the receivers where it constitutes interference while
at the same time it has to remain distinct from these same subspaces
at its corresponding receiver.
\begin{lem}[Sparsity requirement for decoding]
\label{lem:spreq}If $D=\left(1-\epsilon\right)L/2$ and $V_{i},\, i=1,\dots,K$
satisfy the decoding condition at all the receivers, then $\mathrm{sp}_{N}\left(V_{i}\right)\ge2N-\epsilon L$
for all $i$ and $N=1,...,D$.\end{lem}
\begin{IEEEproof}
Assume the contrary that for one of the subspaces $V_{i}$, $\mathrm{sp}_{N}\left(V_{i}\right)<2N-\epsilon L$
for some $N=1,...,D$. This implies that there exists $S\subseteq\left\{ 1,...,L\right\} $
such that $|S|<2N-\epsilon L$ and $\dim\left(V_{i}\cap\mathbb{R}^{S}\right)\ge N$,
and hence $2\dim\left(V_{i}\cap\mathbb{R}^{S}\right)-\left|S\right|>\epsilon L$.
Consider the signal space at receiver $i$, which is $\mathbf{H}_{ii}V_{i}$,
and the interference space from transmitter 1 (assume $i$ is not
1 or 2), which is $\mathbf{H}_{i1}V_{1}$. From the decoding condition
at receiver $i$, we have $\mathbf{H}_{ii}V_{i}\cap\mathbf{H}_{i1}V_{1}=\{0\}$,
or equivalently $V_{1}\cap\mathbf{H}_{i1}^{-1}\mathbf{H}_{ii}V_{i}=\{0\}$.
Note that 
\[
\dim\left(\left(\mathbf{H}_{i1}^{-1}\mathbf{H}_{ii}V_{i}\right)\cap\mathbb{R}^{S}\right)=\dim\left(V_{i}\cap\mathbb{R}^{S}\right)>\left(\left|S\right|+\epsilon L\right)/2,
\]
and since $V_{1}\cap\mathbf{H}_{i1}^{-1}\mathbf{H}_{ii}V_{i}=\{0\}$,
we have 
\[
\dim\left(V_{1}\cap\mathbb{R}^{S}\right)<\left|S\right|-\left(\left|S\right|+\epsilon L\right)/2=\left(\left|S\right|-\epsilon L\right)/2.
\]

Consider the interference at receiver 2, we have 
\[
\dim\left(\mathbf{H}_{21}V_{1}+\mathbf{H}_{2i}V_{i}\right)=\dim\left(V_{1}+\mathbf{H}_{21}^{-1}\mathbf{H}_{2i}V_{i}\right)\le\left(1+\epsilon\right)L/2,
\]
but 
\begin{eqnarray*}
\lefteqn{\dim\left(V_{1}+\mathbf{H}_{21}^{-1}\mathbf{H}_{2i}V_{i}\right)}\\
 & \ge & \dim\left(V_{1}+\left(\left(\mathbf{H}_{21}^{-1}\mathbf{H}_{2i}V_{i}\right)\cap\mathbb{R}^{S}\right)\right)\\
 & > & D+\left(\left|S\right|+\epsilon L\right)/2-\dim\left(V_{1}\cap\left(\left(\mathbf{H}_{21}^{-1}\mathbf{H}_{2i}V_{i}\right)\cap\mathbb{R}^{S}\right)\right)\\
 & > & \left(1-\epsilon\right)L/2+\left(\left|S\right|+\epsilon L\right)/2-\left(\left|S\right|-\epsilon L\right)/2\\
 & = & \left(1+\epsilon\right)L/2,
\end{eqnarray*}
which leads to a contradiction. 
\end{IEEEproof}

\subsection{Linear Independence Condition}

Next, we state a property of the matrices $\mathbf{T}_{ijk}$, which
we need in order to prove our main result. 
\begin{defn}[Linear independence condition]
We say that a set of diagonal matrices $\left\{ \mathbf{T}_{i}\right\} _{i=1,...,M}\subseteq\mathbb{R}^{L\times L}$
with nonzero diagonal entries satisfies the \emph{linear independence
condition} if for any set of integer vectors $A\subseteq\mathbb{Z}^{M}$,
and $\mathbf{v}\in\mathbb{R}^{L}$ with $\left\Vert \mathbf{v}\right\Vert _{0}\ge\left|A\right|$,
the set of vectors 
\[
\left\{ \prod_{i=1}^{M}\mathbf{T}_{i}^{x_{i}}\mathbf{v}\,:\,\mathbf{x}=[x_{1},...,x_{M}]^{T}\in A\right\} 
\]
is linearly independent. 
\end{defn}
Almost all of the sets of diagonal matrices satisfy the linear independence
condition, as shown in the following lemma. 
\begin{lem}
\label{lem:subset-indep}Let $\mathbf{T}_{i}\in\mathbb{R}^{L\times L}$
($i=1,...,M$, $M\ge2$) be diagonal matrices. Consider the $LM$-dimensional
space containing all such $\left\{ \mathbf{T}_{i}\right\} $ with
the Lebesgue measure. Then $\left\{ \mathbf{T}_{i}\right\} $ satisfies
the linear independence condition almost everywhere.\end{lem}
\begin{IEEEproof}
Fix any $A\subseteq\mathbb{Z}^{M}$. It suffices to consider the case
where all entries of $\mathbf{v}$ are nonzero and $\left|A\right|=L$.
Write $\Phi\left(\mathbf{x}\right)=\prod_{i=1}^{M}\mathbf{T}_{i}^{x_{i}}$.
To show $\left\{ \Phi\left(\mathbf{x}\right)\mathbf{v}\,:\,\mathbf{x}\in A\right\} $
is linearly independent for any $\mathbf{v}$ with nonzero entries,
since $\Phi\left(\mathbf{x}\right)$ are diagonal matrices, it suffices
to show that $\left\{ \mathrm{diag}\left(\Phi\left(\mathbf{x}\right)\right)\,:\,\mathbf{x}\in A\right\} $
(the vector formed by diagonal entries) are linearly independent.

Let $A=\left\{ \mathbf{x}_{1},...,\mathbf{x}_{L}\right\} $, $\mathrm{diag}\left(\mathbf{T}_{i}\right)=\left[t_{i1}\,\cdots\, t_{iL}\right]^{T}$.
Note that $\mathrm{det}\left[\mathrm{diag}\left(\Phi\left(\mathbf{x}_{1}\right)\right)\,\cdots\,\mathrm{diag}\left(\Phi\left(\mathbf{x}_{L}\right)\right)\right]$
is a polynomial (possibly with negative exponents) in $\left\{ t_{i\ell}\right\} _{i=1,...,M,\,\ell=1,...,L}$.
The determinant is zero in a set of nonzero measure only if it is
constantly zero.

Let
$y_{1},...,y_{L}\in\mathbb{R}$. Put $t_{i\ell}=y_{\ell}^{\rho_{i}}$
for certain $\rho_{i}\in\mathbb{Z}$ such that $\sum_{i=1}^{M}\rho_{i}x_{ki}$
are distinct for different $k$, where $\mathbf{x}_{k}=[x_{k1},...,x_{kM}]^{T}$.
Then the determinant 
\begin{align*}
& \mathrm{det}\left[\begin{array}{ccc}
\prod_{i=1}^{M}t_{i1}^{x_{1i}} & \cdots & \prod_{i=1}^{M}t_{i1}^{x_{Li}}\\
\vdots &  & \vdots\\
\prod_{i=1}^{M}t_{iL}^{x_{1i}} & \cdots & \prod_{i=1}^{M}t_{iL}^{x_{Li}}
\end{array}\right] \\
& \;\;\;=\mathrm{det}\left[\begin{array}{ccc}
y_{1}^{\sum_{i=1}^{M}\rho_{i}x_{1i}} & \cdots & y_{1}^{\sum_{i=1}^{M}\rho_{i}x_{Li}}\\
\vdots &  & \vdots\\
y_{L}^{\sum_{i=1}^{M}\rho_{i}x_{1i}} & \cdots & y_{L}^{\sum_{i=1}^{M}\rho_{i}x_{Li}}
\end{array}\right]
\end{align*}
is the product of a Vandermonde polynomial and a Schur polynomial
in $y_{1},...,y_{L}$, and is not constantly zero, which can be shown
easily by induction. Therefore the determinant is nonzero almost everywhere.

To argue that the claim holds for all $A\subseteq\mathbb{Z}^{M}$
almost everywhere, note that the number of subsets of $\mathbb{Z}^{M}$
of size not greater than $L$ is countable. The set of $\left\{ t_{i}\right\} $
for which there exist an $A$ such that the claim is false can be
obtained as the union of countably many sets of measure zero, and
thus is of measure zero. 
\end{IEEEproof}

\subsection{The Linear Algebra Problem}

Let us focus on one of the subspaces, say $V=V_{2}\subseteq\mathbb{R}^{L}$
of transmitter 2. For notational simplicity, we write the set 
\[
\left\{ \mathbf{T}_{2jk}\,:\, j,k\in\left\{ 3,...,K\right\} ,\, j\neq k\right\} \qquad\text{ as}\qquad\left\{ \mathbf{T}_{a}\right\} _{a=1,...,M},
\]
where $M=\left(K-2\right)\left(K-3\right)$. Note that each $\mathbf{T}_{a}$
involves one term $\mathbf{H}_{jk}^{-1}$ which is absent in the definition
of other $\mathbf{T}_{a}$'s, therefore when we consider the $LM$-dimensional
space of the diagonal entries of $\left\{ \mathbf{T}_{a}\right\} _{a=1,...,M}$,
the distribution in that space has a joint probability density. Therefore
by Lemma \ref{lem:subset-indep}, we know that the set $\left\{ \mathbf{T}_{a}\right\} _{a=1,...,M}$
satisfies the linear independence condition with probability 1.

In the earlier sections, we have shown that if we want to approach
the maximal degrees of freedom per user by $\epsilon$, then the decoding
conditions at the receivers imply a lower bound on the sparsity of
$V$ (Lemma~\ref{lem:spreq}) and an upper bound on its alignment width under $\left\{ \mathbf{T}_{a}\right\} _{a=1,...,M}$
in terms of $\epsilon$ (Lemma~\ref{lem:wreq}). In order for a subspace $V$ satisfying these
properties to exist the dimension $L$ of the ambient space should
be large enough. Our goal is to derive a lower bound on $L$ in terms
of $\epsilon$. Thus, we have transformed the interference alignment
problem into the following linear algebra problem: \medbreak \emph{Let
$\left\{ \mathbf{T}_{a}\right\} _{a=1,...,M}$ be diagonal matrices
which satisfy the linear independence condition. Assume $V\subseteq\mathbb{R}^{L}$,
with $\dim V=D=\left(1-\epsilon\right)L/2$, satisfies $\mathrm{sp}_{N}\left(V\right)\ge2N-\epsilon L$
for all $N=1,...,D$, and $\Delta_{\mathbf{T}_{a}}V\le2\epsilon L$
for all $a$. Derive a lower bound on $L$ in terms of $\epsilon$
for such $V$ to exist.}

\section{Lower Bound on Channel Diversity}

In this section, we prove Theorem~\ref{thm:prob_quad}. Before providing
a rigorous proof, we first provide a simpler approximate proof which
captures most of the intuition.

\subsection{Proof Intuition\label{sec:ProofIntuition}}

\begin{figure}[t!]
\centering \scalebox{0.25}{\includegraphics{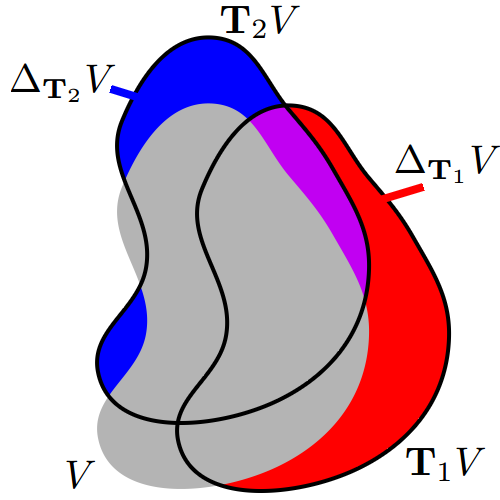}}

\medskip

\scalebox{0.225}{\includegraphics{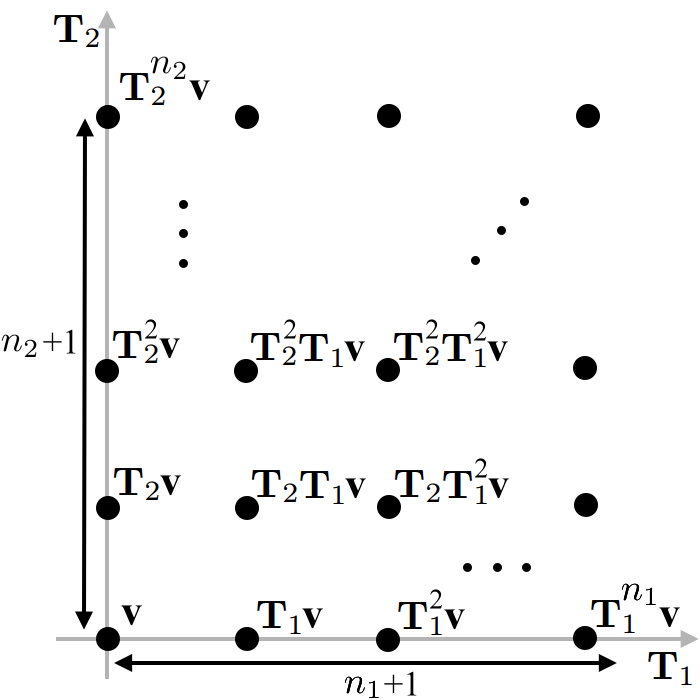}} \caption{\label{fig:gridpt}Top: Illustration of the proof intuition. The area of $V$ ($\dim V$)
cannot be greater than the product of the height ($\Delta_{\mathbf{T}_{1}}V$)
and the width ($\Delta_{\mathbf{T}_{2}}V$). Bottom: Illustration of $W=\mathrm{span}\left\{ \mathbf{T}_{2}^{\alpha_{2}}\mathbf{T}_{1}^{\alpha_{1}}\mathbf{v}\,:\,0\le\alpha_{1}\le n_{1},0\le\alpha_{2}\le n_{2}\right\} $.}

\label{fig:proofcartoon} 
\end{figure}

When $K\ge4$, we have at least two matrices $\mathbf{T}_{1}$ and
$\mathbf{T}_{2}$, and we will use only these two matrices to prove
Theorem \ref{thm:prob_quad}. Recall that $V\subseteq\mathbb{R}^{L}$,
with $\dim V=D=\left(1-\epsilon\right)L/2$, has to have small alignment
width under both of these transformations, i.e., $\Delta_{\mathbf{T}_{1}}V\le2\epsilon L$
and $\Delta_{\mathbf{T}_{2}}V\le2\epsilon L$. In order to get a feel
of the tension these two requirements create, consider Figure~\ref{fig:proofcartoon}.
We can think of $\Delta_{\mathbf{T}_{1}}V$ as relating to the ``length''
of $V$ orthogonal to the ``direction'' $\mathbf{T}_{1}$ and $\Delta_{\mathbf{T}_{2}}V$
as the ``length'' of $V$ orthogonal to the ``direction'' $\mathbf{T}_{2}$.
The area of $V$ ($\dim V$) can not be greater than the product of
the height ($\Delta_{\mathbf{T}_{1}}V$) and the width ($\Delta_{\mathbf{T}_{2}}V$),
therefore 
\[
\dim V=\left(1-\epsilon\right)L/2\approx L/2\leq4\epsilon^{2}L^{2}.
\]
Hence $\epsilon\gtrapprox1/(\sqrt{8L})$. \medbreak 

We next provide an approximate proof which formalizes this intuition.
Before that, we first prove a technical lemma regarding the alignment
width of a subspace. The lemma shows that when we perform successive
extensions (contractions) of a subspace, the dimension of the resultant
subspace increases (decreases) as a concave (convex) function of the
number of extensions (contractions). 
\begin{lem}
\label{lem:width_dec}For any diagonal matrix $\mathbf{T}$ and subspace
$V$, 
\[
\Delta_{\mathbf{T}}\left(\mathrm{e}_{\mathbf{T}}V\right)\le\Delta_{\mathbf{T}}V,
\]
\[
\Delta_{\mathbf{T}}\left(\mathrm{c}_{\mathbf{T}}V\right)\le\Delta_{\mathbf{T}}V.
\]
\end{lem}
\begin{IEEEproof}
Note that 
\begin{align*}
\lefteqn{\Delta_{\mathbf{T}}\left(\mathrm{e}_{\mathbf{T}}V\right)-\Delta_{\mathbf{T}}V}\\
 & =\dim\left(\mathbf{T}^{2}V+\mathbf{T}V+V\right)-2\dim\left(\mathbf{T}V+V\right)+\dim V\\
 & =\dim\left(\mathbf{T}^{2}V+\mathbf{T}V\right)+\dim\left(\mathbf{T}V+V\right)\\
 & \quad-\dim\left(\left(\mathbf{T}^{2}V+\mathbf{T}V\right)\cap\left(\mathbf{T}V+V\right)\right)\\
 & \quad-2\dim\left(\mathbf{T}V+V\right)+\dim V\\
 & =\dim V-\dim\left(\left(\mathbf{T}^{2}V+\mathbf{T}V\right)\cap\left(\mathbf{T}V+V\right)\right)\le0,
\end{align*}
where the second to last line follows from \eqref{eq:dim_sum_int} and the last line follows from $\dim\left(\left(\mathbf{T}^{2}V+\mathbf{T}V\right)\cap\left(\mathbf{T}V+V\right)\right)\geq\dim(\mathbf{T}V)=\dim V$.
A similar result holds for $\Delta_{\mathbf{T}}\left(\mathrm{c}_{\mathbf{T}}V\right)$. 
\end{IEEEproof}
Again when $K\ge4$, we have at least two matrices $\mathbf{T}_{1}$
and $\mathbf{T}_{2}$, and we will use only these two matrices. The
idea of the proof is to find a vector $\mathbf{v}\in V$ and integers
$n_{1},n_{2}$ which are large when $\epsilon$ is small such that
the space 
\begin{align*}
W & =\mathrm{e}_{\mathbf{T}_{2}}^{n_{2}}\mathrm{e}_{\mathbf{T}_{1}}^{n_{1}}\left\langle \mathbf{v}\right\rangle \\
 & =\mathrm{span}\left\{ \mathbf{T}_{2}^{\alpha_{2}}\mathbf{T}_{1}^{\alpha_{1}}\mathbf{v}\,:\,0\le\alpha_{1}\le n_{1},0\le\alpha_{2}\le n_{2}\right\} 
\end{align*}
is a proper subspace of $\mathbb{R}^{L}$. By the linear independence
condition of $\mathbf{T}_{1}$ and $\mathbf{T}_{2}$, we can then
have $(n_{1}+1)(n_{2}+1)<L$ which will allow us to obtain a lower bound
bound for $L$ in terms of epsilon. We can think of $W$ as the span
of the ``grid points'' in the rectangle $\left\{ 0,...,n_{1}\right\} \times\left\{ 0,...,n_{2}\right\} $.
An illustration of the idea is given in Figure~\ref{fig:gridpt}.

\begin{figure}[t!]
\centering \scalebox{0.184}{\includegraphics{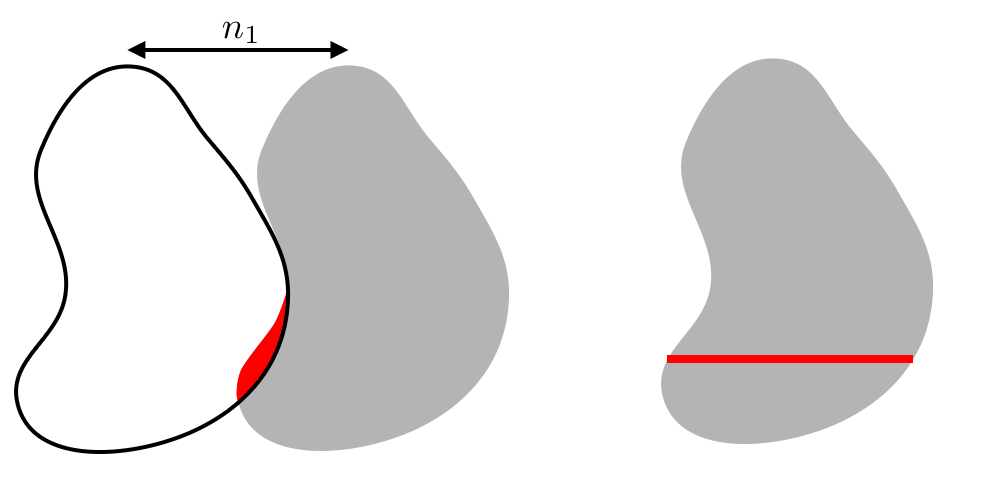}\includegraphics{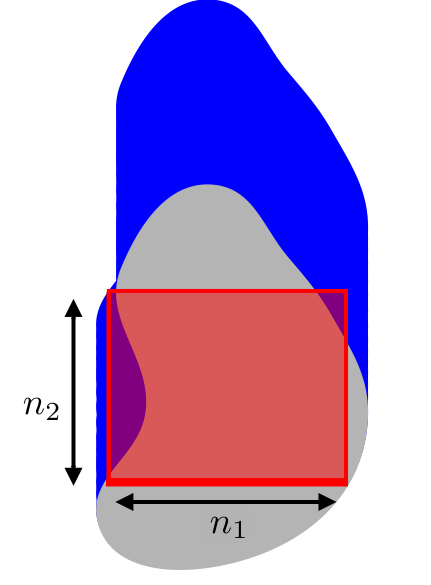}}
\caption{Proof intuition: If $\Delta_{\mathbf{T}_{1}}V$ and $\Delta_{\mathbf{T}_{2}}V$
are both small then a set of vectors $\left\{ \mathbf{T}_{2}^{\alpha_{2}}\mathbf{T}_{1}^{\alpha_{1}}\mathbf{v}\,:\,0\le\alpha_{1}\le n_{1},0\le\alpha_{2}\le n_{2}\right\} $
spans a proper subspace of $\mathbb{R}^{L}$ for some large $n_{1}$
and $n_{2}$.}

\label{fig:proofintuition} 
\end{figure}

We will first find a long ``line'' $\mathrm{e}_{\mathbf{T}_{1}}^{n_{1}}\left\langle \mathbf{v}\right\rangle $
which is a subspace of $V$. %We make use of the small alignment width of $V$ along $\mathbf{T}_{1}$ together with Lemma~\ref{lem:width_dec}. 
Note that if we perform a contraction in $\mathbf{T}_{1}$ direction,
the resultant subspace $\mathrm{c}_{\mathbf{T}_{1}}V$, compared to
$V$, will have dimension reduced by $\Delta_{\mathbf{T}_{1}}V$.
If we perform a second contraction, by Lemma~\ref{lem:width_dec},
the resultant subspace $\mathrm{c}_{\mathbf{T}_{1}}^{2}V$, will have
dimension reduced by at most $\Delta_{\mathbf{T}_{1}}V$ as compared
to $\mathrm{c}_{\mathbf{T}_{1}}V$, therefore at most $2\Delta_{\mathbf{T}_{1}}V$
as compared to $V$. Following in this manner, this means that as
long as $n_{1}\Delta_{\mathbf{T}_{1}}V<\dim V$, the resultant subspace
$\mathrm{c}_{\mathbf{T}_{1}}^{n_{1}}V$ after we perform $n_{1}$
contractions will still be nonempty. Hence we can find 
\[
\widetilde{\mathbf{v}}\in\mathrm{c}_{\mathbf{T}_{1}}^{n_{1}}V=V\cap\mathbf{T}_{1}V\cap\cdots\cap\mathbf{T}_{1}^{n_{1}}V.
\]
This means $\widetilde{\mathbf{v}},\mathbf{T}_{1}^{-1}\widetilde{\mathbf{v}},...,\mathbf{T}_{1}^{-n_{1}}\widetilde{\mathbf{v}}\in V$.
Let $\mathbf{v}=\mathbf{T}_{1}^{-n_{1}}\widetilde{\mathbf{v}}$, then
$\mathrm{e}_{\mathbf{T}_{1}}^{n_{1}}\left\langle \mathbf{v}\right\rangle \subseteq V$.

Next we find $n_{2}$. Again %by the small alignment width of $V$ along $\mathbf{T}_{2}$, 
we know the dimension of $\mathrm{e}_{\mathbf{T}_{2}}V$ is larger
by $\Delta_{\mathbf{T}_{2}}V$ as compared to $V$, and moreover by
Lemma~\ref{lem:width_dec} if we perform multiple extensions the
dimension of the resultant subspace increases by at most $\Delta_{\mathbf{T}_{2}}V$
at each step. Hence, as long as $n_{2}\Delta_{\mathbf{T}_{2}}V<L-\dim V$,
we can perform $n_{2}$ extensions and the resultant subspace $\mathrm{e}_{\mathbf{T}_{2}}^{n_{2}}V$
will still be a proper subspace of $\mathbb{R}^{L}$. Since $W=\mathrm{e}_{\mathbf{T}_{2}}^{n_{1}}\mathrm{e}_{\mathbf{T}_{1}}^{n_{2}}\left\langle \mathbf{v}\right\rangle \subseteq\mathrm{e}_{\mathbf{T}_{2}}^{n_{2}}V$,
$W$ is also a proper subspace of $\mathbb{R}^{L}$.

We finally use the linear independence condition for $\mathbf{T}_{1}$
and $\mathbf{T}_{2}$ to conclude that for any $n_{1}$ and $n_{2}$
such that $n_{1}\Delta_{\mathbf{T}_{1}}V<\dim V$ and $n_{2}\Delta_{\mathbf{T}_{2}}V<L-\dim V$,
$(n_{1}+1)(n_{2}+1)<L$. Now, since $\Delta_{\mathbf{T}_{2}}V,\Delta_{\mathbf{T}_{1}}V\leq2\epsilon L$
and $\dim V=(1-\epsilon)L/2$, we can take any $n_{1}$ and $n_{2}$
such that 
\begin{align*}
n_{1} & <\dim V/\Delta_{\mathbf{T}_{1}}V=(1-\epsilon)/4\epsilon\approx1/4\epsilon,\\
n_{2} & <\left(L-\dim V\right)/\Delta_{\mathbf{T}_{2}}V=(1+\epsilon)/4\epsilon\approx1/4\epsilon,
\end{align*}
which gives the lower bound $L\gtrapprox\epsilon^{-2}/16$ on the
channel diversity $L$ in terms of the gap $\epsilon$ to the optimal
degrees of freedom. Note that the smaller $\epsilon$ we want to achieve,
the larger $L$ we need. Equivalently, $\epsilon\geq1/4\sqrt{L}$.
This proof idea is illustrated pictorially in Figure~\ref{fig:proofintuition}.

A few details are missing in this proof intuition. For example, the
entries of $\mathbf{v}$ may be zero, so $\dim W$ may be smaller
than $(n_{1}+1)(n_{2}+1)$. This is where we need to control the sparsity
of the subspace $V$. A rigorous proof is given in the next subsection.

\subsection{Proof of Theorem~\ref{thm:prob_quad}}

\label{sec:LBDiversity} In this subsection, we give the proof of Theorem
\ref{thm:prob_quad}, which is implied by the following theorem. 
\begin{thm}
\label{thm:Lquad}Let $\mathbf{T}_{1},\mathbf{T}_{2}\in\mathbb{R}^{L\times L}$
satisfy the linear independence condition. Let $\epsilon>0$. Assume
there exist vector subspace $V\subseteq\mathbb{R}^{L}$ with $\dim V=D=\left(1-\epsilon\right)L/2$
satisfying $\mathrm{sp}_{N}\left(V\right)\ge2N-\epsilon L$ for any
$N=1,...,D$, and $\Delta_{\mathbf{T}_{1}}V,\Delta_{\mathbf{T}_{2}}V\le2\epsilon L$,
then we have 
\[
L\ge\epsilon^{-2}/121.
\]
\end{thm}
\begin{IEEEproof}
Note that for any $n_{1}\ge0$, by Lemma \ref{lem:width_dec}, 
\[
\dim\mathrm{c}_{\mathbf{T}_{1}}^{n_{1}}V\ge\dim V-n_{1}\Delta_{\mathbf{T}_{1}}V.
\]
Substitute $n_{1}=\left\lfloor \frac{D-N}{2\epsilon L}\right\rfloor $
for some $N$.  Since $\Delta_{\mathbf{T}_{1}}V\le2\epsilon L$, we have $\dim\mathrm{c}_{\mathbf{T}_{1}}^{n_{1}}V\ge N$,
and therefore since $\mathrm{c}_{\mathbf{T}_{1}}^{n_{1}}V\subseteq V$,
by the definition of sparsity for $V$, we can find $\left\Vert \mathbf{v}\right\Vert _{0}\ge\mathrm{sp}_{N}\left(V\right)$
such that $\mathbf{v}\in\mathbf{T}_{1}^{-n_{1}}\mathrm{c}_{\mathbf{T}_{1}}^{n_{1}}V = \mathrm{c}_{\mathbf{T}_{1}^{-1}}^{n_{1}}V$,
and hence $\mathrm{e}_{\mathbf{T}_{1}}^{n_{1}}\left\langle \mathbf{v}\right\rangle \subseteq V$ by~\eqref{eq:ec_contain_eq}.

On the other hand, for any $n_{2}\ge0$, by Lemma \ref{lem:width_dec},
\[
\dim\mathrm{e}_{\mathbf{T}_{2}}^{n_{2}}V\le\dim V+n_{2}\Delta_{\mathbf{T}_{2}}V.
\]
Substitute $n_{2}=\left\lfloor \frac{\mathrm{sp}_{N}\left(V\right)-1-D}{2\epsilon L}\right\rfloor $, since $\Delta_{\mathbf{T}_{2}}V\le2\epsilon L$,
we have $\dim\mathrm{e}_{\mathbf{T}_{2}}^{n_{2}}V\le\mathrm{sp}_{N}\left(V\right)-1$.
Since $\mathrm{e}_{\mathbf{T}_{2}}^{n_{2}}\mathrm{e}_{\mathbf{T}_{1}}^{n_{1}}\left\langle \mathbf{v}\right\rangle \subseteq\mathrm{e}_{\mathbf{T}_{2}}^{n_{2}}V$,
we also have $\dim\mathrm{e}_{\mathbf{T}_{2}}^{n_{2}}\mathrm{e}_{\mathbf{T}_{1}}^{n_{1}}\left\langle \mathbf{v}\right\rangle \le\mathrm{sp}_{N}\left(V\right)-1$.

Note that by the linear independence condition %for $\mathbf{T}_{1}$ and $\mathbf{T}_{2}$, 
\begin{align*}
\lefteqn{\dim\mathrm{e}_{\mathbf{T}_{1}}^{n_{1}}\mathrm{e}_{\mathbf{T}_{2}}^{n_{2}}\left\langle \mathbf{v}\right\rangle }\\
 & =\dim\mathrm{span}\left\{ \mathbf{T}_{1}^{\alpha_{1}}\mathbf{T}_{2}^{\alpha_{2}}\mathbf{v}\,:\,0\le\alpha_{1}\le n_{1},0\le\alpha_{2}\le n_{2}\right\} \\
 & =\min\left\{ \left(n_{1}+1\right)\left(n_{2}+1\right),\,\left\Vert \mathbf{v}\right\Vert _{0}\right\} .
\end{align*}
Since $\left\Vert \mathbf{v}\right\Vert _{0}\ge\mathrm{sp}_{N}\left(V\right)$
and $\dim\mathrm{e}_{\mathbf{T}_{2}}^{n_{2}}\mathrm{e}_{\mathbf{T}_{1}}^{n_{1}}\left\langle \mathbf{v}\right\rangle \le\mathrm{sp}_{N}\left(V\right)-1$,
$\dim \mathrm{e}_{\mathbf{T}_{2}}^{n_{2}}\mathrm{e}_{\mathbf{T}_{1}}^{n_{1}}\left\langle \mathbf{v}\right\rangle =\left(n_{1}+1\right)\left(n_{2}+1\right)$.
Hence we have 
\begin{eqnarray*}
\mathrm{sp}_{N}\left(V\right)-1 & \ge & \left(n_{1}+1\right)\left(n_{2}+1\right)\\
 & \ge & \left(\frac{D-N}{2\epsilon L}\right)\left(\frac{\mathrm{sp}_{N}\left(V\right)-D}{2\epsilon L}\right),
\end{eqnarray*}
\begin{eqnarray*}
4\epsilon^2 L^2 & \ge & \frac{\left(D-N\right)\left(\mathrm{sp}_{N}\left(V\right)-D\right)}{\mathrm{sp}_{N}\left(V\right)-1}\\
 & \ge & \frac{\left(D-N\right)\left(\mathrm{sp}_{N}\left(V\right)-D\right)}{\mathrm{sp}_{N}\left(V\right)}.
\end{eqnarray*}
Recall that $D=\left(1-\epsilon\right)L/2$. Substitute $N=\left\lceil \frac{3+\epsilon}{8}L\right\rceil $.
Note that $\mathrm{sp}_{N}\left(V\right)\ge2N-\epsilon L\ge\frac{3}{2}D$.
\begin{eqnarray*}
4\epsilon^2 L^2 & \ge & \frac{\left(D-N\right)\left(\mathrm{sp}_{N}\left(V\right)-D\right)}{\mathrm{sp}_{N}\left(V\right)}\\
 & \ge & \frac{1}{3}\left(\frac{1-\epsilon}{2}L-\frac{3+\epsilon}{8}L-1\right)\\
 & = & \frac{1-5\epsilon}{24}L-\frac{1}{3}.
\end{eqnarray*}

Note that $2D=\left(1-\epsilon\right)L<L$, and since both sides are
integers, $2D\le L-1$ and $L\epsilon\ge1$. We split the analysis
into two cases: if $\epsilon\ge1/121$, then $L\ge\epsilon^{-1}\ge\frac{1}{121\epsilon^{2}}$;
if $\epsilon<1/121$, then $L\ge\epsilon^{-1}>121$. If $L\ge122$
then 
\begin{eqnarray*}
4\epsilon^2 L^2 & \ge & \frac{1-5\epsilon}{24}L-\frac{1}{3}\\
 & \ge & \frac{1-5\epsilon}{24}L-\frac{L}{366}\\
 & = & \frac{57-305\epsilon}{1464}L\\
 & \ge & \frac{57-305/121}{1464}L\\
 & = & \frac{824}{22143}L.
\end{eqnarray*}
Hence 
\[
L\ge\frac{206}{22143\epsilon^{2}}\ge\frac{1}{121\epsilon^{2}}.
\]
This completes the proof of the theorem. 
\end{IEEEproof}

\section{Generalization to Block Fading\label{sec:BlockFading}}

%In this section, we consider the block fading case, where the channel
%coefficients are constant over $L$ coherence periods of duration
%$T$, i.e., $\mathbf{H}_{ij}=\mathbf{I}_{T}\otimes\mathrm{diag}\left[h_{ij}^{(1)},...,h_{ij}^{(L)}\right]\in\mathbb{R}^{TL\times TL}$,
%where $\otimes$ denotes the Kronecker product.

In this section, we consider the block fading case, where the channel
coefficients are constant over $L$ coherence periods of duration
$T$. Let $\mathbf{H}_{ij}=\mathrm{diag}\left[h_{ij}^{(1)},...,h_{ij}^{(L)}\right]\otimes \mathbf{I}_{T}\in\mathbb{R}^{TL\times TL}$,
where $\otimes$ denotes the Kronecker product.

Define $\mathbf{P}_{k}\in\mathbb{R}^{T\times TL}$ such that $\left(\mathbf{P}_{k}\right)_{ij}=1$
when $j=T\left(k-1\right)+i$, $\left(\mathbf{P}_{k}\right)_{ij}=0$
otherwise. Note that $\mathbf{P}_{k}$ is the projection which selects
the entries of a vector in $\mathbb{R}^{TL}$ that are in the $k$-th
coherence period.

It is easy to observe that the width requirement for decoding remains
the same in this case, i.e., if $D=\left(1-\epsilon\right)TL/2$ and
$V_{i}\subseteq\mathbb{R}^{TL},\, i=1,\dots,K$ satisfy the decoding
condition at all the receivers, then $\Delta_{\mathbf{T}_{ijk}}V_{i}\le2\epsilon TL$
for all distinct $i,j,k\neq1$ and again focusing on a single subspace
$V=V_{2}$, we have $\Delta_{\mathbf{T}_{i}}V\le2\epsilon TL,\,\,\forall i=1,...,M$
where $M=\left(K-2\right)\left(K-3\right)$.

We will need to generalize the definition of sparsity for $V\subseteq\mathbb{R}^{TL}$
as follows. Let 
\[
\mathrm{sp}^{(T)}\left(V\right)=\sum_{k=1}^{L}\dim\left(\mathbf{P}_{k}V\right),
\]
\begin{eqnarray}
\lefteqn{\mathrm{sp}^{(T)}_{N}\left(V\right)} & \nonumber\\
& \!\!=\!\! & \!\!\min\left\{ \sum_{k=1}^{L}\!\dim\widetilde{W}_{k}:\widetilde{W}_{k}\!\subseteq\!\mathbb{R}^{T}\!,\dim\left(\!V\!\cap\!\sum_{k=1}^{L}\mathbf{P}_{k}^{T}\widetilde{W}_{k}\right)\!\ge\! N\!\right\} \nonumber\\
 & \!\!=\!\! & \!\!\min\left\{ \mathrm{sp}^{(T)}\left(W\right)\,:\, W\subseteq V,\,\dim\left(W\right)\ge N\right\}.\label{def:blockNspar}
\end{eqnarray}

Note that when $T=1$, $\mathrm{sp}^{(T)}\left(W\right)$ counts the
number of positions where the vectors in $W$ have non-zero entries,
i.e., $\mathrm{sp}^{(1)}\left(W\right)=\max_{\mathbf{v}\in W}\left\Vert \mathbf{v}\right\Vert _{0}$,
therefore the new definition of $N$-sparsity coincides with the earlier
one in this case. Note that for larger $T$, we consider the dimension
of each $T$-length portion of $W$, $\dim\left(\mathbf{P}_{k}W\right)$,
instead of simply counting the positions with non-zero entries.

We can observe the following properties for $\mathrm{sp}^{(T)}_{N}\left(V\right)$,
which will be used in the following section: 
\begin{equation}
\mathrm{sp}^{(T)}_{N}\left(\mathrm{c}_{\mathbf{T}}V\right)\ge\mathrm{sp}^{(T)}_{N}\left(V\right),\label{eq:sparprop1}
\end{equation}
\begin{equation}
\mathrm{sp}^{(T)}_{N+\Delta_{\mathbf{T}}V}\left(\mathrm{e}_{\mathbf{T}}V\right)\ge\mathrm{sp}^{(T)}_{N}\left(V\right),\label{eq:sp_exbound}
\end{equation}
for any diagonal matrix $\mathbf{T}\in\mathbb{R}^{TL\times TL}$.
The first inequality simply follows from the fact that $\mathrm{c}_{\mathbf{T}}V\subseteq V$.
The second inequality follows from the fact that if there exists $\widetilde{W}_{k}\subseteq\mathbb{R}^{T}$,
$W=\sum_{k=1}^{L}\mathbf{P}_{k}^{T}\widetilde{W}_{k}$ such that $\dim(\mathrm{e}_{\mathbf{T}}V\cap W)\geq N+\Delta_{\mathbf{T}}V$
then $\dim(V\cap W)\geq N$. Therefore, $\mathrm{sp}^{(T)}_{N+\Delta_{\mathbf{T}}V}\left(\mathrm{e}_{\mathbf{T}}V\right)\ge\mathrm{sp}^{(T)}_{N}\left(V\right)$.

The following lemma is the analogue of Lemma~\ref{lem:spreq} and
establishes the corresponding sparsity requirement for the block fading
case. 
\begin{lem}[Sparsity requirement for decoding]
\label{lem:spreq-1}If $D=\left(1-\epsilon\right)TL/2$ and $V_{i},\, i=1,\dots,K$
satisfy the decoding condition at all the receivers, then $\mathrm{sp}^{(T)}_{N}\left(V_{i}\right)\ge2N-\epsilon TL$
for all $i$ and $N=1,...,D$.\end{lem}
\begin{IEEEproof}
Assume the contrary that $\mathrm{sp}^{(T)}_{N}\left(V_{i}\right)<2N-\epsilon TL$
for some $N=1,...,D$. This implies that there exists $\widetilde{W}_{k}\subseteq\mathbb{R}^{T}$,
$W=\sum_{k=1}^{L}\mathbf{P}_{k}^{T}\widetilde{W}_{k}$ such that $\dim W<2N-\epsilon TL$
and $\dim\left(V_{i}\cap W\right)\ge N$, or equivalently $2\dim\left(V_{i}\cap W\right)-\dim W>\epsilon TL$.
Consider the signal at receiver $i$, which is $\mathbf{H}_{ii}V_{i}$,
and the interference from transmitter 1 (assume $i$ is not 1 or 2),
which is $\mathbf{H}_{i1}V_{1}$. From the decoding condition at receiver
$i$, we have $\mathbf{H}_{ii}V_{i}\cap\mathbf{H}_{i1}V_{1}=\{0\}$,
$V_{1}\cap\mathbf{H}_{i1}^{-1}\mathbf{H}_{ii}V_{i}=\{0\}$. Note
that 
\begin{align*}
\mathbf{H}_{ij}W & =\sum_{k=1}^{L}\mathbf{H}_{ij}\mathbf{P}_{k}^{T}\widetilde{W}_{k}\\
 & =\sum_{k=1}^{L}\left(\left(\mathbf{H}_{ij}\right)_{T(k-1)+1,\,T(k-1)+1}\mathbf{I}\right)\mathbf{P}_{k}^{T}\widetilde{W}_{k}\\
 & =\sum_{k=1}^{L}\mathbf{P}_{k}^{T}\widetilde{W}_{k}\\
 & =W,
\end{align*}
and hence, 
\begin{align*}
\dim\left(\left(\mathbf{H}_{i1}^{-1}\mathbf{H}_{ii}V_{i}\right)\cap W\right) & =\dim\left(V_{i}\cap\left(\mathbf{H}_{i1}\mathbf{H}_{ii}^{-1}W\right)\right)\\
 & =\dim\left(V_{i}\cap W\right)\\
 & >\left(\dim W+\epsilon TL\right)/2.
\end{align*}
Combining this with $V_{1}\cap\mathbf{H}_{i1}^{-1}\mathbf{H}_{ii}V_{i}=\{0\}$,
we have $\dim\left(V_{1}\cap W\right)<\left(\dim W-\epsilon TL\right)/2$.

Consider the interference at receiver 2, we have $\dim\left(\mathbf{H}_{21}V_{1}+\mathbf{H}_{2i}V_{i}\right)=\dim\left(V_{1}+\mathbf{H}_{21}^{-1}\mathbf{H}_{2i}V_{i}\right)\le\left(1+\epsilon\right)TL/2$,
but 
\begin{eqnarray*}
\lefteqn{\dim\left(V_{1}+\mathbf{H}_{21}^{-1}\mathbf{H}_{2i}V_{i}\right)}\\
 & \ge \!\!\!\!\!& \dim\left(V_{1}+\left(\left(\mathbf{H}_{21}^{-1}\mathbf{H}_{2i}V_{i}\right)\cap W\right)\right)\\
 & > \!\!\!\!\!& D\! +\! \left(\dim W+\epsilon TL\right)/2\! -\dim\left(V_{1}\cap\left(\left(\mathbf{H}_{21}^{-1}\mathbf{H}_{2i}V_{i}\right)\cap W\right)\right)\\
 & > \!\!\!\!\!& \left(1-\epsilon\right)TL/2+\left(\dim W+\epsilon TL\right)/2-\left(\dim W-\epsilon TL\right)/2\\
 & = \!\!\!\!\!& \left(1+\epsilon\right)TL/2,
\end{eqnarray*}
which leads to a contradiction. 
\end{IEEEproof}
We next generalize the linear independence condition for diagonal
matrices which we defined in the earlier section to a block linear
independence condition. One can again verify that the new condition
reduces to the linear independence condition in the earlier section when
$T=1$. 
\begin{defn}\label{blockind}
We call a set of diagonal matrices $\left\{ \mathbf{T}_{i}\right\} _{i=1,...,M}\subseteq\mathbb{R}^{TL\times TL}$
with nonzero diagonal entries, where $\mathbf{T}_{i}=\widetilde{\mathbf{T}}_{i}\otimes \mathbf{I}_{T}$,
satisfies the \emph{block linear independence condition} if for any
set of integer vectors $A\subseteq\mathbb{Z}^{M}$ with $\left|A\right|=L$,
and $V\subseteq\mathbb{R}^{TL}$, we have 
\[
\dim\left(\sum_{\mathbf{x}\in A}\left(\left(\prod_{i=1}^{M}\mathbf{T}_{i}^{x_{i}}\right)V\right)\right)=\mathrm{sp}^{(T)}\left(V\right).
\]

\end{defn}
Almost all of the sets of diagonal matrices satisfy the block linear
independence condition, as shown in the following lemma. 
\begin{lem}
\label{lem:subset-indep-1}Let $\mathbf{T}_{i}\in\mathbb{R}^{TL\times TL}$
($i=1,...,M$, $M\ge2$) be diagonal matrices where $\mathbf{T}_{i}=\widetilde{\mathbf{T}}_{i}\otimes \mathbf{I}_{T}$.
Consider the $LM$-dimensional space containing all such $\left\{ \mathbf{T}_{i}\right\} $
with the Lebesgue measure. Then $\left\{ \mathbf{T}_{i}\right\} $
satisfies the block linear independence condition almost everywhere.\end{lem}
\begin{IEEEproof}
Let $A=\left\{ \mathbf{x}_{1},...,\mathbf{x}_{L}\right\} $, and $\widetilde{\Phi}\left(\mathbf{x}\right)=\prod_{i=1}^{M}\widetilde{\mathbf{T}}_{i}^{x_{i}}$.
As shown in Lemma \ref{lem:subset-indep}, the matrix $\mathbf{X}=\left[\mathrm{diag}\left(\widetilde{\Phi}\left(\mathbf{x}_{1}\right)\right)\,\cdots\,\mathrm{diag}\left(\widetilde{\Phi}\left(\mathbf{x}_{L}\right)\right)\right]$
is full rank for any $A$ almost everywhere. Hence 
\begin{align*}
\lefteqn{\sum_{\mathbf{x}\in A}\left(\left(\prod_{i=1}^{M}\mathbf{T}_{i}^{x_{i}}\right)V\right)}& \\
 & =\sum_{i=1}^{L}\left(\widetilde{\Phi}\left(\mathbf{x}_{i}\right) \otimes \mathbf{I}_{T}\right)V\\
 & =\sum_{i=1}^{L}\left(\sum_{j=1}^{L}\left(\mathbf{X}^{-1}\right)_{ij}\left(\widetilde{\Phi}\left(\mathbf{x}_{j}\right) \otimes \mathbf{I}_{T}\right)\right)V\\
 & =\sum_{i=1}^{L}\left(\left(\sum_{j=1}^{L}\left(\mathbf{X}^{-1}\right)_{ij}\widetilde{\Phi}\left(\mathbf{x}_{j}\right)\right)\otimes \mathbf{I}_{T}\right)V\\
 & =\sum_{i=1}^{L}\left(\mathbf{D}_i \otimes \mathbf{I}_{T}\right)V\\
 & =\sum_{i=1}^{L}\mathbf{P}_{i}^{T}\mathbf{P}_{i}V,
\end{align*}
where $\mathbf{D}_{i} \in \mathbb{R}^{L \times L}$ is the diagonal matrix with 1 at the $i$-th position and 0 elsewhere, and $\mathbf{X}^{-1}$ denotes the inverse of the matrix $\mathbf{X}$.
Note that each of $\mathbf{P}_{i}^{T}\mathbf{P}_{i}V$ has disjoint
support, and therefore 
\begin{align*}
\dim\left(\sum_{\mathbf{x}\in A}\left(\left(\prod_{i=1}^{M}\mathbf{T}_{i}^{x_{i}}\right)V\right)\right) & =\dim\left(\sum_{i=1}^{L}\mathbf{P}_{i}^{T}\mathbf{P}_{i}V\right)\\
 & =\sum_{i=1}^{L}\dim\left(\mathbf{P}_{i}^{T}\mathbf{P}_{i}V\right)\\
 & =\mathrm{sp}^{(T)}\left(V\right).
\end{align*}

\end{IEEEproof}
Theorem \ref{thm:prob_quad_block} follows immediately
from the following theorem.
\begin{thm}
\label{thm:Lquad_block}Let $\mathbf{T}_{1},\mathbf{T}_{2}\in\mathbb{R}^{TL\times TL}$
satisfy the block linear independence condition. Let $\epsilon>0$.
Assume there exist vector subspace $V\subseteq\mathbb{R}^{TL}$ with
$\dim V=D=\left(1-\epsilon\right)TL/2$ satisfying $\mathrm{sp}^{(T)}_{N}\left(V\right)\ge2N-\epsilon TL$
for any $N$, and $\Delta_{\mathbf{T}_{1}}V,\Delta_{\mathbf{T}_{2}}V\le2\epsilon TL$,
then we have 
\[
L\ge\epsilon^{-2}/400.
\]
\end{thm}
\begin{IEEEproof}
Note that for any $n_{1}\ge0$, by Lemma \ref{lem:width_dec}, 
\[
\dim\mathrm{c}_{\mathbf{T}_{1}}^{n_{1}}V\ge\dim V-n_{1}\Delta_{\mathbf{T}_{1}}V.
\]
Let $W=\mathbf{T}_{1}^{-n_{1}}\mathrm{c}_{\mathbf{T}_{1}}^{n_{1}}V = \mathrm{c}_{\mathbf{T}_{1}^{-1}}^{n_{1}}V$. Substitute $n_{1}=\left\lfloor \frac{D-N}{2\epsilon TL}\right\rfloor $
for some $N$, we have $\dim W\ge N$, and therefore since $W\subseteq V$,
by the definition of sparsity for $V$, $\mathrm{sp}^{(T)}\left(W\right)\ge\mathrm{sp}^{(T)}_{N}\left(V\right)$.
Also note that $\mathrm{e}_{\mathbf{T}_{1}}^{n_{1}}W\subseteq V$  by~\eqref{eq:ec_contain_eq}.

On the other hand, for any $n_{2}\ge0$, by Lemma \ref{lem:width_dec},
\[
\dim\mathrm{e}_{\mathbf{T}_{2}}^{n_{2}}V\le\dim V+n_{2}\Delta_{\mathbf{T}_{2}}V.
\]
Substitute $n_{2}=\left\lfloor \frac{\mathrm{sp}^{(T)}_{N}\left(V\right)-1-D}{2\epsilon TL}\right\rfloor $,
we have $\dim\mathrm{e}_{\mathbf{T}_{2}}^{n_{2}}V\le\mathrm{sp}^{(T)}_{N}\left(V\right)-1$.
Since $\mathrm{e}_{\mathbf{T}_{1}}^{n_{1}}\mathrm{e}_{\mathbf{T}_{2}}^{n_{2}}W\subseteq\mathrm{e}_{\mathbf{T}_{2}}^{n_{2}}V$,
we also have $\dim\mathrm{e}_{\mathbf{T}_{1}}^{n_{1}}\mathrm{e}_{\mathbf{T}_{2}}^{n_{2}}W\le\mathrm{sp}^{(T)}_{N}\left(V\right)-1$.

Note that by the block linear independence condition, if $\left(n_{1}+1\right)\left(n_{2}+1\right)\ge L$,
then $\dim\mathrm{e}_{\mathbf{T}_{1}}^{n_{1}}\mathrm{e}_{\mathbf{T}_{2}}^{n_{2}}W\ge\mathrm{sp}^{(T)}_{N}\left(V\right)$,
which leads to a contradiction. Hence 
\begin{eqnarray*}
L & > & \left(n_{1}+1\right)\left(n_{2}+1\right)\\
 & \ge & \left(\frac{D-N}{2\epsilon TL}\right)\left(\frac{\mathrm{sp}^{(T)}_{N}\left(V\right)-D}{2\epsilon TL}\right),
\end{eqnarray*}
\begin{eqnarray*}
4\epsilon^2 T^2 L^2 & \ge & L^{-1}\left(D-N\right)\left(\mathrm{sp}^{(T)}_{N}\left(V\right)-D\right).
\end{eqnarray*}
Recall that $D=\left(1-\epsilon\right)TL/2$. Substitute $N=\left\lceil \frac{3-\epsilon}{8}TL\right\rceil $,
by $\mathrm{sp}^{(T)}_{N}\left(V\right)\ge2N-\epsilon TL$, 
\begin{align*}
4\epsilon^2 T^2 L^2 & \ge  L^{-1}\left(\frac{1-\epsilon}{2}TL-\frac{3-\epsilon}{8}TL-1\right)\\
 &\;\;\;\;\;\;\;\cdot\left(\frac{3-\epsilon}{4}TL-\epsilon TL-\frac{1-\epsilon}{2}TL\right)\\
 & =  \frac{1-3\epsilon}{4}T\left(\frac{1-3\epsilon}{8}TL-1\right)\\
 & \ge  \frac{\left(1-3\epsilon\right)^{2}}{32}T^{2}L-\frac{1}{4}T.
\end{align*}
Note that $2D=\left(1-\epsilon\right)TL<TL$, since both sides are
integers, $2D\le TL-1$, $TL\epsilon\ge1$. If $\epsilon\ge1/20$,
then $L\ge1\ge\frac{1}{400\epsilon^{2}}$.

If $\epsilon<1/20$, then $TL\ge\epsilon^{-1}>20$, $TL\ge21$, and
\begin{eqnarray*}
4\epsilon^2 T^2 L^2 & \ge & \frac{\left(1-3\epsilon\right)^{2}}{32}T^{2}L-\frac{1}{4}T\\
 & \ge & \frac{\left(1-3/20\right)^{2}}{32}T^{2}L-\frac{T^{2}L}{84}\\
 & = & \frac{2869}{268800}T^{2}L.
\end{eqnarray*}
Hence 
\[
L\ge\frac{2869}{1075200\epsilon^{2}}\ge\frac{1}{400\epsilon^{2}}.
\]

\end{IEEEproof}

\section{A Tighter Scaling Bound when $K$ grows\label{sec:BetterBound}}

In this section, we prove Theorem \ref{thm:better_bound}, which states
%\[
%\mathrm{DoF}\le\frac{K}{2}\left(1-C\min\left\{ %\frac{1}{\sqrt[4]{L}},\frac{2^{\left(K-2\right)\left(K-3\right)/8}}{\sqrt{L}}\right\} \right),
%\]
\[
\mathrm{DoF}\le\frac{K}{2}\left(1-2^{-17}\min\left\{ \frac{1}{\sqrt[4]{L}},\frac{2^{\left(K-2\right)\left(K-3\right)/4}}{\sqrt{L}}\right\} \right).
\]
%where $C=2^{-18}$.
This implies that the gap decreases at most as
$1/\sqrt[4]{L}$ when $L$ is smaller than the order of $2^{(K-2)(K-3)}$.
%\textcolor{red}{As a result, when $K$ grows, either we need an exponential channel diversity $L = \Omega(2^{\left(K-2\right)\left(K-3\right)})$, or the gap to the optimal 1/2 degrees of freedom per user decreases at most like $1/\sqrt[4]{L}$.}
We only consider
the block fading case in this section, since the fast fading case
can be treated as a special case of block fading with $T=1$.

\begin{figure}[t!]
\centering \scalebox{0.3}{\includegraphics{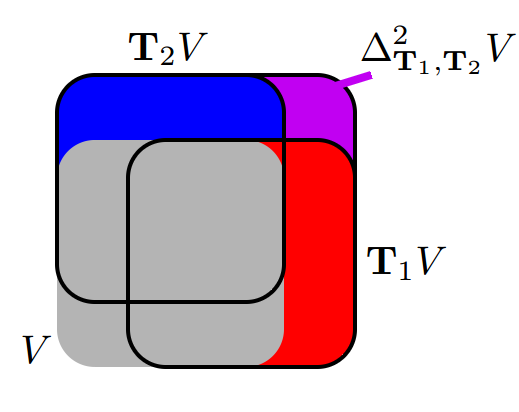}} \caption{Illustration of the second order extension width of $V$ under $\mathbf{T}_{1},\mathbf{T}_{2}$.
Multiplication by $\mathbf{T}_{1}$ and $\mathbf{T}_{2}$ are represented
by a shift to the right and upward respectively.}

\label{fig:alinmentwidth-1} 
\end{figure}

As we have seen in the previous sections, proving that the gap decreases
at most as $1/\sqrt{L}$ only requires to use the alignment width condition for two matrices, $\mathbf{T}_{1}$ and $\mathbf{T}_{2}$, one used for extension and one for contraction,  i.e., we only need $M\ge2$. Proving a gap larger than $1/\sqrt{L}$
requires to use more than one matrix for extension and contraction. We believe the proof of this theorem is important in suggesting one way in which this can be done. We introduce the notions of second order extension and contraction widths, which describe how an extension (contraction)
in $\mathbf{T}_{2}$ would increase (decrease) the alignment width
under $\mathbf{T}_{1}$. This is illustrated in Figure~\ref{fig:alinmentwidth-1}.
\begin{defn}
(Second order extension and contraction width) For a subspace $V$
and diagonal matrices $\mathbf{T}_{1},\mathbf{T}_{2}$, define the
\emph{second order extension width} $\Delta_{\mathbf{T}_{1},\mathbf{T}_{2}}^{2}V$
and \emph{second order contraction width} $\nabla_{\mathbf{T}_{1},\mathbf{T}_{2}}^{2}V$
by 
\begin{align*}
\Delta_{\mathbf{T}_{1},\mathbf{T}_{2}}^{2}V & = \dim\mathrm{e}_{\mathbf{T}_{1}}\mathrm{e}_{\mathbf{T}_{2}}V-\dim\mathrm{e}_{\mathbf{T}_{1}}V-\dim\mathrm{e}_{\mathbf{T}_{2}}V+\dim V\\
 & = \Delta_{\mathbf{T}_{1}}\mathrm{e}_{\mathbf{T}_{2}}V-\Delta_{\mathbf{T}_{1}}V,\\
\nabla_{\mathbf{T}_{1},\mathbf{T}_{2}}^{2}V & = \dim\mathrm{c}_{\mathbf{T}_{1}}\mathrm{c}_{\mathbf{T}_{2}}V-\dim\mathrm{c}_{\mathbf{T}_{1}}V-\dim\mathrm{c}_{\mathbf{T}_{2}}V+\dim V\\
 & = \Delta_{\mathbf{T}_{1}}V-\Delta_{\mathbf{T}_{1}}\mathrm{c}_{\mathbf{T}_{2}}V.
\end{align*}

\end{defn}

Note that the second order extension and contraction widths can be either positive or negative. An important relation is that

\begin{equation}\label{eq:secondorder}
\Delta_{\mathbf{T}_{1},\mathbf{T}_{2}}^{2}V\le\nabla_{\mathbf{T}_{1},\mathbf{T}_{2}}^{2}V,
\end{equation}

which follows from~\eqref{eq:ec_contain} by observing that
\begin{align*}
&\nabla_{\mathbf{T}_{1},\mathbf{T}_{2}}^{2}V-\Delta_{\mathbf{T}_{1},\mathbf{T}_{2}}^{2} \\
&= \left(\Delta_{\mathbf{T}_{1}}V-\Delta_{\mathbf{T}_{1}}\mathrm{c}_{\mathbf{T}_{2}}V \right) - \left(\Delta_{\mathbf{T}_{2}}\mathrm{e}_{\mathbf{T}_{1}}V-\Delta_{\mathbf{T}_{2}}V \right)\\
&= \left(\dim\left(\mathrm{e}_{\mathbf{T}_{1}}V\right)- \dim V -\dim\left(\mathrm{e}_{\mathbf{T}_{1}}\mathrm{c}_{\mathbf{T}_{2}}V\right) + \dim\left(\mathrm{c}_{\mathbf{T}_{2}}V\right) \right)\\
&\;\;\; - \left(\dim\left(\mathrm{e}_{\mathbf{T}_{1}}V\right)- \dim\left(\mathrm{c}_{\mathbf{T}_{2}}\mathrm{e}_{\mathbf{T}_{1}}V\right) -\dim V + \dim\left(\mathrm{c}_{\mathbf{T}_{2}}V\right) \right)\\
&=\dim\left(\mathrm{c}_{\mathbf{T}_{2}}\mathrm{e}_{\mathbf{T}_{1}}V\right)-\dim\left(\mathrm{e}_{\mathbf{T}_{1}}\mathrm{c}_{\mathbf{T}_{2}}V\right) \\
&\ge0.
\end{align*}
This implies that for any number $a\ge 0$, either $a\ge \Delta_{\mathbf{T}_{1},\mathbf{T}_{2}}^{2}V$ (i.e., extension of $V$ by $\mathbf{T}_{2}$ increases $\Delta_{\mathbf{T}_{1}}$ by at most $a$), or $a\le \nabla_{\mathbf{T}_{1},\mathbf{T}_{2}}^{2}V$ (i.e., contraction of $V$ by $\mathbf{T}_{2}$ decreases $\Delta_{\mathbf{T}_{1}}$ by at least $a$) (A respective comment holds when $a\leq 0$.) Intuitively, at least one of contraction or extension by $\mathbf{T}_{2}$ would produce a subspace with small alignment width with respect to $\mathbf{T}_{1}$, i.e. by choosing the respective operation with respect to $\mathbf{T}_{2}$, the new subspace  can be made to either have a $\Delta_{\mathbf{T}_{1}}$ which is not much larger than that of the original subspace or even smaller than that.

In Theorem~\ref{thm:Lquad}, we only use one matrix for extension, and another for contraction.
To prove the stronger Theorem \ref{thm:better_bound}, we use all the matrices
$\mathbf{T}_{1},...,\mathbf{T}_{M}$ (recall $M=\left(K-2\right)\left(K-3\right)$). We consider the average alignment
width of a subspace under these matrices. The main idea of the proof
is that, we perform extension and contraction repeatedly on the subspace
$V$. In each step, we keep the average alignment width small, which guarantees
that there exist a matrix among $\mathbf{T}_{1},...,\mathbf{T}_{M}$
with a small alignment width. This matrix is then used for the extension
or contraction in the next step.
\begin{defn}
(Average alignment width) Define $\overline{\Delta}V=\frac{1}{M}\sum_{j=1}^{M}\Delta_{\mathbf{T}_{j}}V$
to be the \emph{average alignment width} of subspace $V$ along all $\mathbf{T}_{i}$'s,
and similarly define 
\begin{eqnarray*}
\overline{\Delta}_{\mathbf{T}}^{2}V & = & \frac{1}{M}\sum_{j=1}^{M}\Delta_{\mathbf{T},\mathbf{T}_{j}}^{2}V,\\
\overline{\nabla}_{\mathbf{T}}^{2}V & = & \frac{1}{M}\sum_{j=1}^{M}\nabla_{\mathbf{T},\mathbf{T}_{j}}^{2}V.
\end{eqnarray*}
\end{defn}
By the property of second order alignment width in \eqref{eq:secondorder}, $\overline{\Delta}_{\mathbf{T}}^{2}V\le\overline{\nabla}_{\mathbf{T}}^{2}V$.
Intuitively, at least one of contraction or extension by $\mathbf{T}$ would produce a subspace with small average alignment width.

As seen in the proof of Theorem~\ref{thm:Lquad}, our goal is to perform as many extensions on $V$ as possible such that the resultant subspace has dimension less than that of the whole space, and to perform as many contractions on $V$ as possible such that the dimension is greater than 0. The number of consecutive extensions/contractions performed directly affects the bound on $L$. This number is in turn dictated by the alignment width of the subspace, which is the increase in dimension after performing an extension (and the decrease in dimension after performing a contraction), and therefore it determines how many further extensions/contractions can be performed.

We will next present several lemmas which are useful in proving Theorem
\ref{thm:better_bound}. The following lemma shows that we can either
perform extension on a subspace repeatedly to obtain a subspace with similar dimension, sparsity and average alignment width, or find another subspace
with smaller average alignment width and similar dimension and sparsity.

The intuition behind this lemma is that we can perform extensions on $W$ using different matrices (unlike Theorem~\ref{thm:Lquad} which uses the same matrix repeatedly) to obtain
$\mathrm{e}_{\mathbf{T}_{k_{1}}}\cdots\mathrm{e}_{\mathbf{T}_{k_{\tilde{n}-1}}}W$,
until the next extension $\mathrm{e}_{\mathbf{T}_{k_{1}}}\cdots\mathrm{e}_{\mathbf{T}_{k_{\tilde{n}}}}W$
would increase the average alignment width too much. If such event does not happen, then we obtain a long series of extensions such that the resultant subspace does not have an average alignment width much larger than the original one. If such an event happens (i.e. the extension $\mathrm{e}_{\mathbf{T}_{k_{\tilde{n}}}}$ increases the average alignment width too much), by the property of second order alignment width in \eqref{eq:secondorder}, we know that the contraction $\mathrm{c}_{\mathbf{T}_{k_{\tilde{n}}}}$ can be used to significantly decrease the average alignment width. By performing the contraction instead of extension, we break the series of extensions but the average alignment width can now be made smaller than what we started with.

\begin{lem}
\label{lem:rep_extend}Let $\mathbf{T}_{j}\in\mathbb{R}^{TL\times TL}$
($j=1,...,M$, $M\ge2$) be diagonal matrices satisfying the block
linear independence condition. For any vector subspace $W\subseteq\mathbb{R}^{TL}$,
subset $S\subseteq\{1,...,M\}$, and strictly increasing sequence
of real numbers $\overline{\Delta}W<a_{1}<\cdots<a_{n}$, $n\ge0$ (assume
$a_{0}=\overline{\Delta}W$), if
\begin{equation}
n\le\left|S\right|-M/2,\label{eq:lemgen_nbound}
\end{equation}
then there exist subspace $\widetilde{W}\subseteq\mathbb{R}^{TL}$ and
$\widetilde{n}\in\{0,...,n\}$ such that
\begin{equation}
\left|\dim\widetilde{W}-\dim W\right|\le\delta,\label{eq:lemgen_dim}
\end{equation}
\begin{equation}
\mathrm{sp}^{(T)}_{\, N+\delta}\left(\widetilde{W}\right)\ge\mathrm{sp}^{(T)}_{N}\left(W\right)\label{eq:lemgen_sp}
\end{equation}
for any $N\ge0$, where
\[
\delta=2\sum_{i=0}^{\widetilde{n}-1}a_{i},
\]
and at least one of the following cases holds:
\begin{enumerate}
\item We have $\widetilde{n}\ge1$, and
\begin{equation}
\overline{\Delta}\widetilde{W}\le2a_{\widetilde{n}-1}-a_{\widetilde{n}}.\label{eq:lemgen_ec}
\end{equation}

\item We have $\widetilde{n}=n$,
\begin{equation}
\overline{\Delta}\widetilde{W}\le a_{n},\label{eq:lemgen_ee_w}
\end{equation}
and there exist distinct $k_{1},...,k_{n}\in S$ such that
\begin{equation}
\widetilde{W}=\mathrm{e}_{\mathbf{T}_{k_{1}}}\cdots\mathrm{e}_{\mathbf{T}_{k_{n}}}W.\label{eq:lemgen_ee_eq}
\end{equation}

\end{enumerate}
The same lemma also holds when $\widetilde{W}=\mathrm{e}_{\mathbf{T}_{k_{1}}}\cdots\mathrm{e}_{\mathbf{T}_{k_{n}}}W$
is replaced by $\widetilde{W}=\mathrm{c}_{\mathbf{T}_{k_{1}}}\cdots\mathrm{c}_{\mathbf{T}_{k_{n}}}W$.
We call the former the extension version of the lemma, and the latter
the contraction version.\end{lem}
\begin{IEEEproof}
We prove the extension version $\widetilde{W}=\mathrm{e}_{\mathbf{T}_{k_{1}}}\cdots\mathrm{e}_{\mathbf{T}_{k_{n}}}W$
here. The contraction version is similar. We prove the lemma by induction
on $n$. Note that when $n=0$, we have $\delta=0$, and $\widetilde{W}=W$,
$\widetilde{n}=0$ obviously satisfies \eqref{eq:lemgen_dim}, \eqref{eq:lemgen_sp},
\eqref{eq:lemgen_ee_w} and \eqref{eq:lemgen_ee_eq}. We then consider
$n\ge1$ and assume the lemma is true for $n-1$.

By Markov inequality, we have 
\begin{equation}
\left|\left\{ j\in\{1,...,M\}\,:\,\Delta_{\mathbf{T}_{j}}W\le2\overline{\Delta}W\right\} \right|\ge M/2.\label{eq:twice_w}
\end{equation}
By \eqref{eq:lemgen_nbound}, we have $\left|S\right|>M/2$, hence
we can always find $k\in S$ such that $\Delta_{\mathbf{T}_{k}}W\le2\overline{\Delta}W$.
Consider two cases:

\bigskip{}

\noindent\textbf{Case 1:} $\overline{\Delta}_{\mathbf{T}_{k}}^{2}W>a_{1}-\overline{\Delta}W$,

We will check that $\widetilde{W}=\mathrm{c}_{\mathbf{T}_{k}}W$ and $\widetilde{n}=1$
satisfies \eqref{eq:lemgen_dim}, \eqref{eq:lemgen_sp} and \eqref{eq:lemgen_ec}. First we bound $\Delta_{\mathbf{T}_{k}}W$ by
\[
\Delta_{\mathbf{T}_{k}}W \le 2\overline{\Delta}W =2a_{0} = 2\sum_{i=0}^{\widetilde{n}-1}a_{i} =\delta.
\]
%\begin{align*}
% \Delta_{\mathbf{T}_{k}}W & \le 2\overline{\Delta}W\\
% & =2a_{0}\\
% & =\delta.
%\end{align*}

For \eqref{eq:lemgen_dim}, $\dim\widetilde{W}\le\dim W$, and
\begin{align*}
\dim\widetilde{W} & \ge\dim W-\Delta_{\mathbf{T}_{k}}W\\
 & \ge\dim W-\delta.
\end{align*}

Note that \eqref{eq:lemgen_sp} follows directly from \eqref{eq:sp_exbound} and $\Delta_{\mathbf{T}_{k}}W \le \delta$.
Finally for \eqref{eq:lemgen_ec},
\begin{align*}
\overline{\Delta}\widetilde{W} & =\overline{\Delta}W-\overline{\nabla}_{\mathbf{T}_{k}}^{2}W\\
 & \le\overline{\Delta}W-\overline{\Delta}_{\mathbf{T}_{k}}^{2}W\\
 & <\overline{\Delta}W-\left(a_{1}-\overline{\Delta}W\right)\\
 & =2a_{0}-a_{1}.
\end{align*}

\bigskip{}

\noindent\textbf{Case 2:} $\overline{\Delta}_{\mathbf{T}_{k}}^{2}W\le a_{1}-\overline{\Delta}W$,

Let $\widehat{W}=\mathrm{e}_{\mathbf{T}_{k}}W$. We apply induction hypothesis
on the subspace $\widehat{W}$, subset $S\backslash\left\{ k\right\} $
and sequence $a_{2},...,a_{n}$. To check \eqref{eq:lemgen_nbound},
\begin{align*}
n-1 & \le\left|S\right|-1-M/2\\
 & =\left|S\backslash\left\{ k\right\} \right|-M/2.
\end{align*}
Hence there exist $\widetilde{W}$ satisfying \eqref{eq:lemgen_dim},
\eqref{eq:lemgen_sp}, and either \eqref{eq:lemgen_ec}, or both \eqref{eq:lemgen_ee_w}
and \eqref{eq:lemgen_ee_eq} with $\widehat{W}$, $S\backslash\left\{ k\right\} $
and $a_{2},...,a_{n}$. We prove that $\widetilde{W}$ satisfies the requirements
for $W$, $S$ and $a_{1},...,a_{n}$ as well.

For \eqref{eq:lemgen_dim},
\begin{align*}
\left|\dim\widetilde{W}\!-\dim W\right|\! & \le\left|\dim\widetilde{W}-\dim\widehat{W}\right|+\left|\dim\widehat{W}-\dim W\right|\\
 & \le2\left(\overline{\Delta}\widehat{W}+\sum_{i=2}^{\widetilde{n}-1}a_{i}\right)+\Delta_{\mathbf{T}_{k}}W\\
 & =2\left(\overline{\Delta}W+\overline{\Delta}_{\mathbf{T}_{k}}^{2}W+\sum_{i=2}^{\widetilde{n}-1}a_{i}\right)+\Delta_{\mathbf{T}_{k}}W\\
 & \le2\left(\overline{\Delta}W+a_{1}-\overline{\Delta}W+\sum_{i=2}^{\widetilde{n}-1}a_{i}\right)+2\overline{\Delta}W\\
 & =2\sum_{i=0}^{\widetilde{n}-1}a_{i}.
\end{align*}

For \eqref{eq:lemgen_sp},
\begin{align*}
\mathrm{sp}^{(T)}_{\, N+2\sum_{i=0}^{\widetilde{n}-1}a_{i}}\left(\widetilde{W}\right) & \ge\mathrm{sp}^{(T)}_{N+2\sum_{i=0}^{\widetilde{n}-1}a_{i}-2\left(\overline{\Delta}\widehat{W}+\sum_{i=2}^{\widetilde{n}-1}a_{i}\right)}\left(\widehat{W}\right)\\
 & =\mathrm{sp}^{(T)}_{N+2a_{1}-2\overline{\Delta}\widehat{W}}\left(\widehat{W}\right)\\
 & =\mathrm{sp}^{(T)}_{N+2a_{1}-2\overline{\Delta}W+2\overline{\Delta}_{\mathbf{T}_{k}}^{2}W}\left(\widehat{W}\right)\\
 & \ge\mathrm{sp}^{(T)}_{N}\left(\widehat{W}\right).
\end{align*}

If \eqref{eq:lemgen_ec} is satisfied for $\widehat{W}$, $S\backslash\left\{ k\right\} $
and $a_{2},...,a_{n}$, then it is clearly also satisfied for $W$,
$S$ and $a_{1},...,a_{n}$ by incrementing $\widetilde{n}$ by one.

If \eqref{eq:lemgen_ee_w} and \eqref{eq:lemgen_ee_eq} are satisfied
for $\widehat{W}$, $S\backslash\left\{ k\right\} $ and $a_{2},...,a_{n}$,
then \eqref{eq:lemgen_ee_w} is clearly also satisfied for $W$, $S$
and $a_{1},...,a_{n}$ since $a_{n}$ is the same in both cases. Also
\eqref{eq:lemgen_ee_eq} directly follows from $\widehat{W}=\mathrm{e}_{\mathbf{T}_{k}}W$.

The result follows from induction.
\end{IEEEproof}
Next we utilize Lemma \ref{lem:rep_extend} repeatedly to show that
given a subspace $W$, there exist two subspaces $\widetilde{W}_{1}$
and $\widetilde{W}_{2}$ with similar dimension, sparsity and average alignment width, such that
the repeated contraction of one of them contains the other one. The
main idea is to apply both the extension and contraction versions
of Lemma \ref{lem:rep_extend} on $W$. If both versions give a repeated
extension and contraction, then those repeated extension and contraction
would satisfy the requirement. Otherwise if one of the versions gives
a subspace with smaller average alignment width, then we can consider
that subspace instead and repeat the process.
\begin{lem}
\label{lem:rep_rep_extend}Let $\mathbf{T}_{j}\in\mathbb{R}^{TL\times TL}$
($j=1,...,M$, $M\ge2$) be diagonal matrices satisfying the block
linear independence condition. For any vector subspace $W\subseteq\mathbb{R}^{TL}$
and integer $n\in\mathbb{Z}_{\ge0}$, $n\le M/4$, there exist subspaces
$\widetilde{W}_{1},\widetilde{W}_{2}\subseteq\mathbb{R}^{TL}$ such that
\begin{equation}
\overline{\Delta}\widetilde{W}_{i}\le2\overline{\Delta}W,\label{eq:lemrr_w}
\end{equation}
\begin{equation}
\left|\dim\widetilde{W}_{i}-\dim W\right|\le\delta,\label{eq:lemrr_dim}
\end{equation}
\begin{equation}
\mathrm{sp}^{(T)}_{\, N+\delta}\left(\widetilde{W}_{i}\right)\ge\mathrm{sp}^{(T)}_{N}\left(W\right)\label{eq:lemrr_sp}
\end{equation}
for any $i\in\{1,2\}$ and $N\ge0$, where
\[
\delta=2^{n+3}\overline{\Delta}W
\]
and there exist distinct $k_{1},...,k_{2n}\in\{1,...,M\}$ such that
\begin{equation}
\widetilde{W}_{1}\subseteq\mathrm{c}_{\mathbf{T}_{k_{1}}}\cdots\mathrm{c}_{\mathbf{T}_{k_{2n}}}\widetilde{W}_{2}.\label{eq:lemrr_eq}
\end{equation}
\end{lem}
\begin{IEEEproof}
We perform induction on $M\overline{\Delta}W$, which is a nonnegative
integer. When $\overline{\Delta}W=0$, then $\Delta_{\mathbf{T}_{k}}W=0$
for all $k$. It can be checked easily that $\widetilde{W}_{1}=\widetilde{W}_{2}=W$
satisfies the conditions. Next we assume the lemma is true for all
subspaces with average alignment width less than $\overline{\Delta}W$ and show that it holds for $W$ with average alignment width $\overline{\Delta}W$.

We invoke Lemma \ref{lem:rep_extend} (extension version) on $W$,
$\{1,...,M\}$ and sequence

\[
a_{i}=\overline{\Delta}W\cdot\left(1+2^{-(n+1)}\left(2^{i+1}-i-2\right)\right),\,\, i=1,...,n.
\]
Note that $a_{i}<2\overline{\Delta}W$. Suppose the lemma gives $\widetilde{W}\subseteq\mathbb{R}^{TL}$
and $\widetilde{n}\in\{0,...,n\}$, which satisfy
\[
\left|\dim\widetilde{W}-\dim W\right|\le2\sum_{i=0}^{\widetilde{n}-1}a_{i}<4\widetilde{n}\overline{\Delta}W,
\]
\[
\mathrm{sp}^{(T)}_{\, N+4\widetilde{n}\overline{\Delta}W}\left(\widetilde{W}\right)\ge\mathrm{sp}^{(T)}_{N}\left(W\right)
\]
for any $N\ge0$. Consider two cases of the outcome of the lemma:

\bigskip{}

\noindent\textbf{Case 1:} $\widetilde{n}\ge1$ and $\overline{\Delta}\widetilde{W}\le2a_{\widetilde{n}-1}-a_{\widetilde{n}},$

Note that
\begin{align*}
\overline{\Delta}\widetilde{W} & \le2a_{\widetilde{n}-1}-a_{\widetilde{n}}\\
 & =\overline{\Delta}W\cdot\bigg(2\left(1+2^{-(n+1)}\left(2^{\widetilde{n}}-\widetilde{n}-1\right)\right) \\
 &\;\;\;\;\;\;\;\;\;-\left(1+2^{-(n+1)}\left(2^{\widetilde{n}+1}-\widetilde{n}-2\right)\right)\bigg)\\
 & =\overline{\Delta}W\cdot\left(1-2^{-(n+1)}\widetilde{n}\right).
\end{align*}
Hence $\overline{\Delta}\widetilde{W}<\overline{\Delta}W$. By applying the induction
hypothesis on $\widetilde{W}$, we obtain $\widetilde{W}_{1}$ and $\widetilde{W}_{2}$.
We will check that they satisfy the conditions.

For \eqref{eq:lemrr_w},
\[
\overline{\Delta}\widetilde{W}_{i}\le2\overline{\Delta}\widetilde{W}<2\overline{\Delta}W.
\]

For \eqref{eq:lemrr_dim},
\begin{align*}
\left|\dim\widetilde{W}_{i}\!-\dim W\right|\! & \le\left|\dim\widetilde{W}_{i}-\dim\widetilde{W}\right|+\left|\dim\widetilde{W}-\dim W\right|\\
 & \le 2^{n+3}\overline{\Delta}\widetilde{W}+4\widetilde{n}\overline{\Delta}W\\
 & \le2^{n+3}\overline{\Delta}W\cdot\left(1-2^{-(n+1)}\widetilde{n}\right)+4\widetilde{n}\overline{\Delta}W\\
 & =2^{n+3}\overline{\Delta}W.
\end{align*}

For \eqref{eq:lemrr_sp},
\begin{align*}
\mathrm{sp}^{(T)}_{\, N+\delta}\left(\widetilde{W}_{i}\right) & = \mathrm{sp}^{(T)}_{\, N+2^{n+3}\overline{\Delta}W}\left(\widetilde{W}_{i}\right) \\
 & \ge \mathrm{sp}^{(T)}_{\, N+2^{n+3}\overline{\Delta}W - 2^{n+3}\overline{\Delta}\widetilde{W}}\left(\widetilde{W}\right)\\
& \ge \mathrm{sp}^{(T)}_{\, N+4\widetilde{n}\overline{\Delta}W}\left(\widetilde{W}\right)\\
 & \ge \mathrm{sp}^{(T)}_{N}\left(W\right).
\end{align*}
Note that \eqref{eq:lemrr_eq}
is satisfied by induction hypothesis.

\bigskip{}

\noindent\textbf{Case 2:} $\widetilde{n}=n$, $\overline{\Delta}\widetilde{W}\le a_{n},$
and there exist distinct $k_{1},...,k_{n}\in S$ such that $\widetilde{W}=\mathrm{e}_{\mathbf{T}_{k_{1}}}\cdots\mathrm{e}_{\mathbf{T}_{k_{n}}}W$,

We invoke Lemma \ref{lem:rep_extend} again, but use the contraction
version instead, on $W$, subset $\{1,...,M\}\backslash\{k_{1},...,k_{n}\}$
and the same sequence $a_{1},...,a_{n}$. To check \eqref{eq:lemgen_nbound},
\begin{align*}
n & \le M/4\\
 & \le\left|\{1,...,M\}\backslash\{k_{1},...,k_{n}\}\right|-M/2.
\end{align*}
Suppose the lemma gives $\widetilde{W}'\subseteq\mathbb{R}^{TL}$ and
$\widetilde{n}'\in\{0,...,n\}$, which satisfy
\[
\left|\dim\widetilde{W}'-\dim W\right|<4\widetilde{n}'\overline{\Delta}W,
\]
\[
\mathrm{sp}^{(T)}_{\, N+4\widetilde{n}'\overline{\Delta}W}\left(\widetilde{W}'\right)\ge\mathrm{sp}^{(T)}_{N}\left(W\right)
\]
for any $N\ge0$.

If the first case of Lemma \ref{lem:rep_extend} holds, then we can
show that $\widetilde{W}'$ satisfies the conditions by the same arguments
as in case~1. Hence we assume the second case holds, that is, $\widetilde{n}'=n$,
$\overline{\Delta}\widetilde{W}'\le a_{n},$ and there exist distinct $k_{n+1},...,k_{2n}\in S$
such that $\widetilde{W}'=\mathrm{c}_{\mathbf{T}_{k_{n+1}}}\cdots\mathrm{c}_{\mathbf{T}_{k_{2n}}}W$.
We now check that $\widetilde{W}_{1}=\widetilde{W}'$, $\widetilde{W}_{2}=\widetilde{W}$
satisfies the conditions.

For \eqref{eq:lemrr_w},
\begin{align*}
\overline{\Delta}\widetilde{W}_{i} & \le a_{n}\\
 & =\overline{\Delta}W\cdot\left(1+2^{-(n+1)}\left(2^{n+1}-n-2\right)\right)\\
 & <2\overline{\Delta}W.
\end{align*}

For \eqref{eq:lemrr_dim},
\begin{align*}
\left|\dim\widetilde{W}_{i}-\dim W\right| & \le4n\overline{\Delta}W\\
 & \le2^{n+3}\overline{\Delta}W.
\end{align*}

Similar holds for \eqref{eq:lemrr_sp}. For \eqref{eq:lemrr_eq},
\begin{align*}
\widetilde{W}_{1} & =\mathrm{c}_{\mathbf{T}_{k_{n+1}}}\cdots\mathrm{c}_{\mathbf{T}_{k_{2n}}}W\\
 & \subseteq\mathrm{c}_{\mathbf{T}_{k_{n+1}}}\cdots\mathrm{c}_{\mathbf{T}_{k_{2n}}}\left(\mathrm{c}_{\mathbf{T}_{k_{1}}}\cdots\mathrm{c}_{\mathbf{T}_{k_{n}}}\widetilde{W}_{2}\right)\\
 & =\mathrm{c}_{\mathbf{T}_{k_{1}}}\cdots\mathrm{c}_{\mathbf{T}_{k_{2n}}}\widetilde{W}_{2}.
\end{align*}

This completes the proof of Lemma \ref{lem:rep_rep_extend}.
\end{IEEEproof}
Next we present a lemma which uses the resultant subspaces of Lemma
\ref{lem:rep_rep_extend} to establish a bound on $L$. It is proved
in a way similar to Theorem \ref{thm:Lquad} and Theorem \ref{thm:Lquad_block}.
\begin{lem}
\label{lem:halfstep}Let $\mathbf{T}_{j}\in\mathbb{R}^{TL\times TL}$
($j=1,...,M$, $M\ge2$) be diagonal matrices satisfying the block
linear independence condition. Let $W_{1},W_{2}\subseteq\mathbb{R}^{TL}$
be subspaces with $\dim W_{1}=D_{1}$, $\dim W_{2}=D_{2}$ satisfying
$W_{1}\subseteq\mathrm{c}_{\mathbf{T}_{3}}\mathrm{c}_{\mathbf{T}_{4}}\cdots\mathrm{c}_{\mathbf{T}_{M}}W_{2}$,
$\mathrm{sp}^{(T)}_{N}\left(W_{1}\right)\ge2N-\alpha$ for any $N$, where
$\alpha>0$, $D_{1}-D_{2}/2-\alpha/2\ge4$, and $\Delta_{\mathbf{T}_{1}}W_{1} ,\, \Delta_{\mathbf{T}_{2}}W_{2} \le 8\epsilon TL$. Then we have 
\[
L^3 \ge 2^{M-10} \epsilon^{-2} T^{-2}\left(D_{1}-D_{2}/2-\alpha/2\right)^{2}.
\]
\end{lem}
\begin{IEEEproof}
Recall that for any $n_{1}\ge0$, 
\[
\dim\mathrm{c}_{\mathbf{T}_{1}}^{n_{1}}W_{1}\ge D_{1}-n_{1}\Delta_{\mathbf{T}_{1}}W_{1}.
\]
Substitute $n_{1}=\left\lfloor \frac{D_{1}-N}{8\epsilon TL}\right\rfloor $
for some $N$, we have $\dim\mathrm{c}_{\mathbf{T}_{1}}^{n_{1}}W_{1}\ge N$.
Let
\begin{align*}
\widehat{W} & = \mathbf{T}_{1}^{-n_{1}} \mathbf{T}_{3}^{-1} \cdots \mathbf{T}_{M}^{-1} \mathrm{c}_{\mathbf{T}_{1}}^{n_{1}}W_{1},
\end{align*}
then
\begin{align*} & \mathrm{e}_{\mathbf{T}_{1}}^{n_{1}}\mathrm{e}_{\mathbf{T}_{3}}\mathrm{e}_{\mathbf{T}_{4}}\cdots\mathrm{e}_{\mathbf{T}_{M}}\widehat{W} \\
& =  \mathrm{e}_{\mathbf{T}_{3}}\mathrm{e}_{\mathbf{T}_{4}}\cdots\mathrm{e}_{\mathbf{T}_{M}}  \mathbf{T}_{3}^{-1} \cdots \mathbf{T}_{M}^{-1} \mathrm{e}_{\mathbf{T}_{1}}^{n_{1}} \mathbf{T}_{1}^{-n_{1}} \mathrm{c}_{\mathbf{T}_{1}}^{n_{1}}W_{1} \\
 & \subseteq  \mathrm{e}_{\mathbf{T}_{3}}\mathrm{e}_{\mathbf{T}_{4}}\cdots\mathrm{e}_{\mathbf{T}_{M}}  \mathbf{T}_{3}^{-1} \cdots \mathbf{T}_{M}^{-1} W_{1} \\
 &\subseteq  \mathrm{e}_{\mathbf{T}_{3}}\mathrm{e}_{\mathbf{T}_{4}}\cdots\mathrm{e}_{\mathbf{T}_{M}}  \mathbf{T}_{3}^{-1} \cdots \mathbf{T}_{M}^{-1} \mathrm{c}_{\mathbf{T}_{3}}\mathrm{c}_{\mathbf{T}_{4}}\cdots\mathrm{c}_{\mathbf{T}_{M}}W_{2}\\
 &\subseteq  W_{2},
\end{align*}
which follows from the fact that the extension and contraction operations commute among themselves and also with multiplication with diagonal matrices and applying the fact that $\mathbf{T}^{-1}\mathrm{e}_{\mathbf{T}}\mathrm{c}_{\mathbf{T}}W \subseteq W$.

On the other hand, for any $n_{2}\ge0$, 
\[
\dim\mathrm{e}_{\mathbf{T}_{2}}^{n_{2}}W_{2}\le D_{2}+n_{2}\Delta_{\mathbf{T}_{2}}W_{2}.
\]
Substitute $n_{2}=\left\lfloor \frac{\mathrm{sp}^{(T)}_{N}\left(W_{1}\right)-1-D_{2}}{8\epsilon TL}\right\rfloor $,
we have $\dim\mathrm{e}_{\mathbf{T}_{2}}^{n_{2}}W_{2}\le\mathrm{sp}^{(T)}_{N}\left(W_{1}\right)-1$.
Let 
\[
\widetilde{W}=\mathrm{e}_{\mathbf{T}_{1}}^{n_{1}}\mathrm{e}_{\mathbf{T}_{2}}^{n_{2}}\mathrm{e}_{\mathbf{T}_{3}}\mathrm{e}_{\mathbf{T}_{4}}\cdots\mathrm{e}_{\mathbf{T}_{M}}\widehat{W}.
\]
Since $\widetilde{W}\subseteq\mathrm{e}_{\mathbf{T}_{2}}^{n_{2}}W_{2}$,
we also have $\dim\widetilde{W}\le\mathrm{sp}^{(T)}_{N}\left(W_{1}\right)-1$.

By the fact that $\mathbf{T}_{j}$ satisfy the block linear independence condition in Definition~\ref{blockind}, if $2^{M-2}\left(n_{1}+1\right)\left(n_{2}+1\right) \ge L$, then
\[
\dim\mathrm{e}_{\mathbf{T}_{1}}^{n_{1}}\mathrm{e}_{\mathbf{T}_{2}}^{n_{2}}\mathrm{e}_{\mathbf{T}_{3}}\mathrm{e}_{\mathbf{T}_{4}}\cdots\mathrm{e}_{\mathbf{T}_{M}}\widehat{W} \,=\, \mathrm{sp}^{(T)}(\widehat{W}).
\]
Combining this with the fact that $\dim\widetilde{W}\le\mathrm{sp}^{(T)}_{N}\left(W_{1}\right)-1$ and that $\mathrm{sp}^{(T)}(\widehat{W}) \,\ge\, \mathrm{sp}^{(T)}_{N}\left(W_{1}\right)$, which follows from the definition of $N$-sparsity in \eqref{def:blockNspar} combined with $\dim \widehat{W} \ge N$ and $\widehat{W} \subseteq \mathbf{T}_{1}^{-n_{1}} \mathbf{T}_{3}^{-1} \cdots \mathbf{T}_{M}^{-1} W_{1}$, which leads to a contradiction,
\begin{align*}
\mathrm{sp}^{(T)}_{N}\left(W_{1}\right)-1 & \ge\, \dim\mathrm{e}_{\mathbf{T}_{1}}^{n_{1}}\mathrm{e}_{\mathbf{T}_{2}}^{n_{2}}\mathrm{e}_{\mathbf{T}_{3}}\mathrm{e}_{\mathbf{T}_{4}}\cdots\mathrm{e}_{\mathbf{T}_{M}}\widehat{W} \\
& \ge\, \mathrm{sp}^{(T)}_{N}(\widehat{W}) \,\ge\, \mathrm{sp}^{(T)}_{N}\left(W_{1}\right).
\end{align*}
Hence,
\begin{eqnarray*}
L & > & 2^{M-2}\left(n_{1}+1\right)\left(n_{2}+1\right)\\
 & \ge & 2^{M-2}\left(\frac{D_{1}-N}{8\epsilon TL}\right)\left(\frac{\mathrm{sp}^{(T)}_{N}\left(W_{1}\right)-D_{2}}{8\epsilon TL}\right),
\end{eqnarray*}
\begin{eqnarray*}
\epsilon^2 T^2 L^3 & \ge & 2^{M-8}\left(D_{1}-N\right)\left(\mathrm{sp}^{(T)}_{N}\left(W_{1}\right)-D_{2}\right).
\end{eqnarray*}
Substitute $N=\biggl\lceil D_{1}/2+D_{2}/4+\alpha/4\biggr\rceil$.
By $\mathrm{sp}^{(T)}_{N}\left(W_{1}\right)\ge2N-\alpha$, we have 
\begin{align*}
\epsilon^2 T^2 L^3 & \ge 2^{M-9}\left(D_{1}-D_{2}/2-\alpha/2-2\right)\left(D_{1}-D_{2}/2-\alpha/2\right)\\
 & \ge 2^{M-10}\left(D_{1}-D_{2}/2-\alpha/2\right)^{2},
\end{align*}
since $D_{1}-D_{2}/2-\alpha/2\ge4$. 
\end{IEEEproof}
Theorem \ref{thm:better_bound} follows
directly from the following theorem.
\begin{thm}
\label{thm:Lexpalt}Let $\mathbf{T}_{j}\in\mathbb{R}^{TL\times TL}$
($j=1,...,M$, $M\ge2$) be diagonal matrices satisfying the block
linear independence condition. If there exist a vector subspace $V\subseteq\mathbb{R}^{TL}$
with $\dim V=\left(1-\epsilon\right)TL/2$ satisfying $\mathrm{sp}^{(T)}_{N}\left(V\right)\ge2N-\epsilon TL$
for any $N$, and $\Delta_{T_{j}}V\le2\epsilon TL$ for any $j$,
then we have 
\[
L\ge2^{-34}\epsilon^{-2}\min\left\{ 2^{M/2},\,\epsilon^{-2}\right\} ,
\]
\end{thm}
\begin{IEEEproof}
If $\epsilon>1/512$, then by Theorem \ref{thm:Lquad_block},
\begin{align*}
L & \ge\epsilon^{-2}/400\\
 & \ge2^{-34}\epsilon^{-4}\\
 & \ge2^{-34}\epsilon^{-2}\min\left\{ 2^{M/2},\,\epsilon^{-2}\right\} .
\end{align*}
Hence we assume $\epsilon\le1/512$ throughout the proof. Note that
since $\dim V=(1-\epsilon)TL/2<TL/2$, we have $(1-\epsilon)TL/2\le(TL-1)/2$,
and therefore $TL\ge\epsilon^{-1}\ge512$.

If $M<8$, then by Theorem \ref{thm:Lquad_block}, 
\begin{align*}
L & \ge\epsilon^{-2}/400\\
 & \ge2^{-34}\epsilon^{-2}\cdot2^{M/2}\\
 & \ge2^{-34}\epsilon^{-2}\min\left\{ 2^{M/2},\,\epsilon^{-2}\right\} .
\end{align*}
Hence we assume $M\ge8$ throughout the proof.

Note that $\overline{\Delta}V\le2\epsilon TL$. We next apply Lemma \ref{lem:rep_rep_extend}
on $V$ by choosing 
\[
n=\min\left\{ \left\lfloor M/4\right\rfloor -1,\,\left\lfloor \log_{2}(\epsilon^{-1})\right\rfloor -9\right\} ,
\]
(note that $n\ge0$ since $M\ge8$ and $\epsilon^{-1}\ge512$). Lemma \ref{lem:rep_rep_extend} guarantees the existences of two subspaces $\widetilde{W}_{1},\widetilde{W}_{2}\subseteq\mathbb{R}^{TL}$
such that $\overline{\Delta}\widetilde{W}_{i}\le4\epsilon TL$, and for any $i\in\{1,2\}$ and $N\ge0$,
\[
\left|\dim\widetilde{W}_{i}-\left(1-\epsilon\right)TL/2\right|\le2^{n+4}\epsilon TL,
\]
\[
\mathrm{sp}^{(T)}_{N+2^{n+4}\epsilon TL}\left(\widetilde{W}_{i}\right)\ge \mathrm{sp}^{(T)}_{N}(V), 
\]
where using $\mathrm{sp}^{(T)}_{N}\left(V\right)\ge2N-\epsilon TL$, the last inequality implies
\[
\mathrm{sp}^{(T)}_{N}\left(\widetilde{W}_{i}\right)\ge2N-\left(2^{n+5}+1\right)\epsilon TL.
\]
By Lemma \ref{lem:rep_rep_extend}, there also exist distinct $k_{3},...,k_{2n+2}\in\{1,...,M\}$
such that
\begin{equation}
\widetilde{W}_{1}\subseteq\mathrm{c}_{\mathbf{T}_{k_{3}}}\cdots\mathrm{c}_{\mathbf{T}_{k_{2n+2}}}\widetilde{W}_{2}.\label{eq:lemrr_eq-1}
\end{equation}
Since $M-2n-1>M/2$, by applying the same argument as in \eqref{eq:twice_w}
twice, we can find $k_{1},k_{2}$ such that $k_{1},k_{2},k_{3},...,k_{2n+2}$
are distinct, $\Delta_{\mathbf{T}_{k_{1}}}\widetilde{W}_{1},\,\Delta_{\mathbf{T}_{k_{2}}}\widetilde{W}_{2}\le8\epsilon TL$.

We apply Lemma \ref{lem:halfstep} on $\widetilde{W}_{1}$, $\widetilde{W}_{2}$
and $\mathbf{T}_{k_{1}},....,\mathbf{T}_{k_{2n+2}}$. Let $\alpha=\left(2^{n+5}+1\right)\epsilon TL$. We first check that the condition $\dim\widetilde{W}_{1}-\frac{1}{2}\dim\widetilde{W}_{2}-\frac{\alpha}{2}\geq 4$ required in Lemma \ref{lem:halfstep} is satisfied:
\begin{align*}
 & \dim\widetilde{W}_{1}-\frac{1}{2}\dim\widetilde{W}_{2}-\frac{1}{2}\left(2^{n+5}+1\right)\epsilon TL\\
 & =\left(\dim\widetilde{W}_{1}-\frac{\left(1-\epsilon\right)TL}{2}\right)-\frac{1}{2}\left(\dim\widetilde{W}_{2}-\frac{\left(1-\epsilon\right)TL}{2}\right) \\
 &\;\;\;\;\;\;\;+\frac{\left(1-\epsilon\right)TL}{4}-\frac{1}{2}\left(2^{n+5}+1\right)\epsilon TL\\
 & \ge-2^{n+4}\epsilon TL-2^{n+3}\epsilon TL+\frac{1}{4}TL-\left(2^{n+4}+\frac{3}{4}\right)\epsilon TL\\
 & =\frac{1}{4}TL-\left(\frac{5}{4}\cdot 2^{n+5}+\frac{3}{4}\right)\epsilon TL\\
 & \ge\frac{1}{4}TL-2^{n+6}\epsilon TL\\
 & \ge\frac{1}{4}TL-2^{\log(\epsilon^{-1})-9+6}\epsilon TL\\
 & =\frac{1}{8}TL\\
 & \ge64
\end{align*}
since $TL\ge\epsilon^{-1}\ge512$. Hence Lemma \ref{lem:halfstep}
gives
\[
 L^3\ge 2^{2n-8} \epsilon^{-2} T^{-2}\! \left(\!\dim\widetilde{W}_{1}\!-\!\frac{1}{2}\dim\widetilde{W}_{2}\!-\!\frac{1}{2}\!\left(2^{n+4}\!+\!1\right)\!\epsilon TL\!\right)^{2}\!\!\!,
\]
\[
L^3\ge 2^{2n-8} \epsilon^{-2} T^{-2}\left(\frac{1}{8}TL\right)^{2},
\]
By taking the logarithm of both sides, we obtain
\begin{align*}
\log_{2}L & \ge2n-14+2\log_{2}(\epsilon^{-1})\\
 & \ge2\min\left\{ M/4-2,\,\log(\epsilon^{-1})-10\right\} +2\log_{2}(\epsilon^{-1})-14\\
 & \ge\min\left\{ M/2,\,2\log_{2}(\epsilon^{-1})\right\} +2\log_{2}(\epsilon^{-1})-34.
\end{align*}
The result follows.
\end{IEEEproof}
$ $

\section{Conclusion}

In this paper, we derived upper bounds on the degrees of freedom achievable
with vector space interference alignment strategies over the $K$-user
interference channel as a function of the available channel diversity
(the number of independently fading parallel channels). Our results
show that the channel diversity poses a fundamental limit on the efficiency
of interference alignment. In particular, while the gap to the optimal
degrees of freedom is known to decrease inversely proportional to
$L$ for $K=3$, we show that when $K\geq4$ it decreases at most
as $1/\sqrt{L}$. To the best of our knowledge this is the first result
capturing the impact of channel diversity on the achievable degrees
of freedom for $K\geq4$.
%When $K$ is larger than the the order of $\sqrt{\log L}$,
In the regime when $L$ is smaller than
the order of $2^{\left(K-2\right)\left(K-3\right)}$,
we show that the speed of convergence is smaller than $1/\sqrt[4]{L}$.
%\textcolor{red}{When $K$ grows, either we need an exponential channel diversity $L = \Omega(2^{\left(K-2\right)\left(K-3\right)})$, or the speed of convergence is slower than $1/\sqrt[4]{L}$.}
However, there is still a large gap between
the upper bounds we derive and the achievable strategies in the literature, even in the
scaling sense. For example, for $K=4$ the achievability results in
the literature approach the optimal degrees of freedom as $1/\sqrt[5]{L}$
which is significantly slower than $1/\sqrt{L}$. Closing this gap
remains an important problem which will determine the
promise of interference alignment strategies in practical systems. We believe one of the most important contributions of the current paper is to introduce a language (tools and notions) to tackle the problem, which we believe can be further developed to obtain tighter results.

\section*{Acknowledgment}

The authors would like to thank Akshay Venkatesh for the insightful discussions and suggestions.

\bibliographystyle{IEEEtran}
\bibliography{ref,nit}

\newcommand{\noopsort}[1]{}
\begin{thebibliography}{10}
\providecommand{\url}[1]{#1}
\csname url@samestyle\endcsname
\providecommand{\newblock}{\relax}
\providecommand{\bibinfo}[2]{#2}
\providecommand{\BIBentrySTDinterwordspacing}{\spaceskip=0pt\relax}
\providecommand{\BIBentryALTinterwordstretchfactor}{4}
\providecommand{\BIBentryALTinterwordspacing}{\spaceskip=\fontdimen2\font plus
\BIBentryALTinterwordstretchfactor\fontdimen3\font minus
  \fontdimen4\font\relax}
\providecommand{\BIBforeignlanguage}[2]{{%
\expandafter\ifx\csname l@#1\endcsname\relax
\typeout{** WARNING: IEEEtran.bst: No hyphenation pattern has been}%
\typeout{** loaded for the language `#1'. Using the pattern for}%
\typeout{** the default language instead.}%
\else
\language=\csname l@#1\endcsname
\fi
#2}}
\providecommand{\BIBdecl}{\relax}
\BIBdecl

\bibitem{maddahali2008}
M.~Maddah-Ali, A.~Motahari, and A.~Khandani, ``Communication over {MIMO} {X}
  channels: Interference alignment, decomposition, and performance analysis,''
  \emph{IEEE Trans. Info. Theory}, vol.~54, no.~8, pp. 3457--3470, Aug 2008.

\bibitem{Cadambe--Jafar2008}
V.~Cadambe and S.~Jafar, ``Interference alignment and degrees of freedom of the
  {$K$}-user interference channel,'' \emph{{IEEE} Trans. Inf. Theory}, vol.~54,
  no.~8, pp. 3425--3441, Aug. 2008.

\bibitem{OzgurTse08}
A.~\"{O}zg\"{u}r and D.~Tse, ``Achieving linear scaling with interference
  alignment,'' in \emph{Proc. IEEE Int. Symp. on Inform. Theory}, 2009, pp.
  1754 -- 1758.

\bibitem{JafarNow}
\BIBentryALTinterwordspacing
S.~Jafar, ``Interference alignment -- a new look at signal dimensions in a
  communication network,'' \emph{Foundations and Trends in Communications and
  Information Theory}, vol.~7, no.~1, pp. 1--134, 2011. [Online]. Available:
  \url{http://dx.doi.org/10.1561/0100000047}
\BIBentrySTDinterwordspacing

\bibitem{bresler2009}
G.~Bresler and D.~Tse, ``3 user interference channel: Degrees of freedom as a
  function of channel diversity,'' in \emph{Proc. 47th Ann. Allerton Conf.
  Commun., Contr., and Comput.}, 2009, pp. 265--271.

\bibitem{CJW08}
V.~Cadambe, S.~Jafar, and C.~Wang, ``Interference alignment with asymmetric
  complex signaling -settling the host-madsen-nosratinia conjecture,''
  \emph{IEEE Trans. Inform. Theory}, vol.~56, no.~9, pp. 4552--4565, 2010.

\bibitem{li2014channel}
C.~T. Li and A.~{\"O}zg{\"u}r, ``Channel diversity needed for vector
  interference alignment,'' in \emph{Proc. IEEE Int. Symp. Inf. Theory}, 2014,
  pp. 1211--1215.

\bibitem{sun2015interference}
R.~Sun and Z.-Q. Luo, ``Interference alignment using finite and dependent
  channel extensions: The single beam case,'' \emph{IEEE Trans. Inf. Theory},
  vol.~61, no.~1, pp. 239--255, 2015.

\bibitem{gomadam2011}
K.~Gomadam, V.~Cadambe, and S.~Jafar, ``A distributed numerical approach to
  interference alignment and applications to wireless interference networks,''
  \emph{IEEE Trans. Info. Theory}, vol.~57, no.~6, pp. 3309--3322, June 2011.

\bibitem{Yetis}
C.~Yetis, T.~Gou, S.~Jafar, and A.~Kayran, ``On feasibility of interference
  alignment in {MIMO} interference networks,'' \emph{IEEE Trans. Signal
  Processing}, vol.~58, no.~9, pp. 4552--4565, 2010.

\bibitem{bresler2011}
G.~Bresler, D.~Cartwright, and D.~Tse, ``Geometry of the 3-user {MIMO}
  interference channel,'' in \emph{Proc. 49th Ann. Allerton Conf. Commun.,
  Contr., and Comput.}\hskip 1em plus 0.5em minus 0.4em\relax IEEE, 2011, pp.
  1264--1271.

\bibitem{Razaviyayn}
M.~Razaviyayn, G.~Lyubeznik, and Z.~Luo, ``On the degrees of freedom achievable
  through interference alignment in a {MIMO} interference channel,'' \emph{IEEE
  Trans. Signal Processing}, vol.~60, no.~2, pp. 812--821, 2012.

\bibitem{bresler2014}
G.~Bresler, D.~Cartwright, and D.~Tse, ``Feasibility of interference alignment
  for the {MIMO} interference channel,'' \emph{IEEE Trans. Info. Theory},
  vol.~60, no.~9, pp. 5573--5586, Sept 2014.

\end{thebibliography}

%\begin{IEEEbiographynophoto}{Cheuk Ting Li}
%(S'12) received the B.Sc. degree in mathematics and B.Eng. degree in information engineering from The Chinese University of Hong Kong in 2012, and the M.S. degree in electrical engineering from Stanford University in 2014. He is currently working toward the Ph.D. degree at Stanford University. His research interests include information theory and wireless communications.
%\begin{IEEEbiographynophoto}{Ayfer \"{O}zg\"{u}r} (M'06) received her B.Sc. degrees in electrical engineering and physics from Middle East Technical University, Turkey, in 2001 and the M.Sc. degree in communications from the same university in 2004. From 2001 to 2004, she worked as hardware engineer for the Defense Industries Development Institute in Turkey. She received her Ph.D. degree in 2009 from the Information Processing Group at EPFL, Switzerland. In 2010 and 2011, she was a post-doctoral scholar with the Algorithmic Research in Network Information Group at EPFL. She is currently an Assistant Professor in the Electrical Engineering Department at Stanford University. Her research interests include network communications, wireless systems, and information and coding theory. Dr. \"{O}zg\"{u}r received the EPFL Best Ph.D. Thesis Award in 2010 and a NSF CAREER award in 2013. \end{IEEEbiographynophoto}

\end{document}